\documentclass[traditabstract]{aa}

\usepackage{xcolor}%%%%I ADDED THIS ONE MYSELF

\usepackage{graphicx}
\usepackage{txfonts}
\usepackage{comment}%%for big comments

\usepackage{ulem}

\begin{document} 

   \title{Classifying the full SDSS-IV MaNGA Survey using optical diagnostic diagrams: presentation of AGN catalogs in flexible apertures}
   \titlerunning{BPT-selected AGN in MaNGA}

   %\subtitle{I. Overviewing the $\kappa$-mechanism}

   \author{M. Alb\'an \inst{1}
          \and
          D. Wylezalek\inst{1}%{2}%\fnmsep\thanks{Just to show the usage
          %of the elements in the author field}
          }

   \institute{Zentrum f\"ur Astronomie der Universit\"at Heidelberg, Astronomisches Rechen-Institut, M\"onchhofstr, 12-14 69120 Heidelberg, Germany\\
   %for #2 
             }

   \date{Received November 11, 2022; accepted February 16, 2023}

  \abstract{
Accurate active galactic nucleus (AGN) identifications in large galaxy samples are crucial to assess the role of AGN and AGN feedback in the coevolution of galaxies and their central supermassive black holes. Emission line flux ratio diagnostics are the most common technique for identifying AGN in optical spectra. New large samples of integral field unit observations allow The exploration of the role of aperture size used for the classification. In this paper, we present galaxy classifications for all 10,010 galaxies observed within the Mapping Nearby Galaxies at Apache Point Observatory (MaNGA) survey. We use Baldwin-Philips-Terlevich line flux ratio diagnostics combined with an H$\alpha$ equivalent threshold in 60 apertures of varying size for the classification and provide the corresponding catalogs. MaNGA-selected AGN primarily lie below the main sequence of star-forming galaxies, reside with massive galaxies with stellar masses of $\sim 10^{11}$~M$_{\odot}$ and a median H$\alpha$-derived star formation rate of $\sim 1.44$M$_{\odot}$~yr$^{-1}$. We find that the number of `fake' AGN increases significantly beyond selection apertures of $>$~1.0~R$_{eff}$ due to increased contamination from  diffuse ionized gas (DIG). A comparison with previous works shows that the treatment of the underlying stellar continuum and flux measurements can significantly impact galaxy classification. Our work provides the community with AGN catalogs and galaxy classifications for the full MaNGA.% sample derived in a self-consistent way. 
   }


\maketitle
%
%-------------------------------------------------------------------

\section{Introduction}
\label{sec:introduction}
   Mounting observational evidence has shown that supermassive black holes (SMBHs) are ubiquitous in most, or even all, centers of massive galaxies \citep{Decarli2007,Kormendy2013,Graham2016}. Many properties of these galaxies show tight relations \citep[e.g., the bulge's velocity dispersion,][]{Ferrarese2000,Gultekin2009}, in particular with the mass of their SMBHs \citep{Beifiori2011}, suggesting that the presence of a SMBH impacts the evolution of their host galaxies and vice versa. How this happens over cosmic time is an active field of study \citep[e.g.][]{Heckman2014,Park2015,Volonteri2016,Buchner2019,Smith2019,Singh2021}. Standard evolution models inkove the active growth phases of SMBH, i.e. their active galactic nucleus \citep[AGN,][]{Rees1984,Alexander2012,Padovani2017} phase. 
   
   During the SMBH activity, the injection of energy into the interstellar medium (ISM) generated by accretion onto the BH can have a relevant impact on the host galaxy, leading, for example, to quench star formation \citep[negative feedback,][]{Springel2005,Cano-Diaz2012}, enhance it \citep[positive feedback,][]{Mahoro2017,Nesvadba2020,bessiere2022spatially}, or even both \citep[e.g.][]{Wagner2016,Dugan2017}. This has motivated cosmological simulations \citep[see ][for a detailed review]{Somerville2015} to include AGN feedback \citep{Fabian2012} in their models.

   To constrain and accurately quantify the effect that AGN have in their host galaxies, AGN selection algorithms are required to be as complete as possible. AGN galaxies span a very wide peculiar observational characteristics \citep{Maoz_2007}, and it is therefore not surprising that all AGN selection methods come with some caveats. Over the past decades, multiple AGN selection methods have been invoked inspired by the peculiar multi-wavelength radiation from the energetic release of the accretion onto SMBHs  \citep[for a convenient summary of these methods, check section 1.4 from][]{Harrison2014}. However, these techniques can be heavily biased towards, for example, finding AGN preferentially in luminous host galaxies or may not be sensitive to find obscured AGN \citep[e.g.,][]{Azadi2017,yi2022sdss}.
   
   A commonly used AGN selection method is based on strong optical emission-line ratios \citep{baldwin1981,Veilleux1987}, which are frequently referred to as BPT diagrams in the literature. We also adopt that nomenclature in this paper. The used line complexes are close in wavelength space which reduces the effects of dust reddening that could affect the measurement of the emission line ratios \citep{Kennicutt1992}. This technique, together with empirical demarcation lines \citep[for the demarcation line equations and a summary of this methods, see][]{kewley2006}, allows distinguishing between different ionisation sources. For example, star-forming galaxies and HII regions exhibit specific line ratios that tend to occupy a well-defined area in the BPT diagrams. Other celestial objects, such as planetary nebula, and most importantly (for our study, to identify AGN activity) Seyfert and LINER \citep[low ionisation nuclear emission-line region,][]{Halpern1983} galaxies gather in different positions on the BPT diagrams. A frequently used BPT diagnostic diagram includes [O~III]$\lambda$5008/H$\beta$ versus [N~II]$\lambda$6583/H$\alpha$ that distinguishes between AGN-like galaxies, star-forming galaxies, and an overlap region classifying objects as composite galaxies \citep[which show characteristics consistent with both AGN and star-forming galaxies,][]{kewley2001}. However, this diagram does not allow one to distinguish between LINER and Seyfert galaxies by itself.
   
  LINER galaxies have been originally proposed to be driven by weak accretion-powered AGN \citep{Halpern1983,Kauffmann2003,Ho_2008,Masegosa2011}, presenting lower ionization levels \citep{Ferland1983} compared to Seyferts. Therefore, Additional BPT diagrams use [O III]/H$\beta$ versus [S~II] $\lambda\lambda$6717,6731/H$\alpha$ \citep{kewley2001} and [O~III]/H$\beta$ versus [O~I] $\lambda$6302/H$\alpha$ \citep{Kauffmann2003} to further distinguish between LINER and Seyfert galaxies. However, some LINERs have also been reported to be compatible with galaxies whose spectral contribution is dominated by ionizing photons from post asymptotic giant branch stars \citep[e.g.][]{Binette1994,Yan2012,Singh2013} or to be related to poststarburst galaxies \citep{Taniguchi2000}. This situation has motivated alternative diagnostics to select true AGN from the LINER galaxy population. For example, \citep{fernandes2010} proposed to use an additional cut in H$\alpha$ equivalent width (EW) motivated by the fact that Seyfert galaxies have higher EW(H$\alpha$) values than LINER galaxies without AGN.
   
   Some difficulties in using BPT diagnostics \citep[see section 5 of][for a review]{kewley2019} also arise due to the fact that the emission line ratios can be affected by shocks \citep[e.g.,][]{Dopita2002,Allen2008}, obscuration by dust, metallicity \citep[as the BPT diagram correlates with metallicity;][]{Groves2004}, diffuse ionized gas regions \citep{Zhang2016,Mannucci2021}, morphology, and cosmic-time \citep{kewley2013,Hirschmann2017}. Furthermore, observational effects, such as using different aperture sizes affect measurements of integrated galaxy properties, such as integrated star formation rates, emission line fluxes or EW(H$\alpha$) measurements \citep{Hopkins2003,Gomez2003,kewley2005,Iglesias2016}. Naturally, galaxy classification using BPT diagnostics is therefore highly dependent on aperture size \citep{Maragkoudakis2014}. If the galaxies to be classified span a range in redshifts, a constant aperture probes different physical sizes in the observed galaxies. This directly affects the observed emission-line flux ratios in single fiber spectral observations, possibly leading to miss-classifications \citep{Veilleux1995,Maragkoudakis2014}.
   
   Integral field spectroscopy (IFS) \citep{2010Bershady,Drory2015} can help mitigate this effect by allowing to map the 2D spectral properties of a target \citep[e.g.][]{MangaDAP}. In particular, The SDSS-IV survey Mapping Nearby Galaxies at APO (MaNGA) provides optical IFU observations of 10,010 galaxies at $0.03 < z < 0.1$. The final data release DR17 has just recently been released to the public \citep[DR17,][]{abdurrouf2021seventeenth}. In this paper, we investigate the impact of aperture effects for BPT-based AGN classifications as well as provide a suite of AGN catalogs based on three sets of apertures with differing units (kpc, effective radius, and arcsecond). We classify galaxies as star-forming, Seyfert, Composite, LINER, or Ambiguous. 

   The paper is organized as follows: In Section \ref{sec2} we describe the data and some available AGN catalogs from early and recent MaNGA product launches and data releases. The details for the study of the sample and the AGN selection algorithm are described in Section \ref{sec3}. In Section \ref{sec_results} we present our aperture based catalog and discuss the impact of the aperture selection. We compare our AGN candidates with other AGN catalogs in Section \ref{sec_comparisons}. Lastly, we present our conclusions in Section \ref{conclusions}. Throughout the paper we use H$_{0}=$72 km s$^{-1}$ Mpc$^{-1}$, $\Omega_{\rm{M}}$=0.3 and $\Omega_{\Lambda}=0.7$.

\section{Data and AGN catalogs}
\label{sec2}
%
%                                                One column figure
%----------------------------------------------------------------- 
  \begin{figure*}%[h!]
  \includegraphics[width=\hsize]{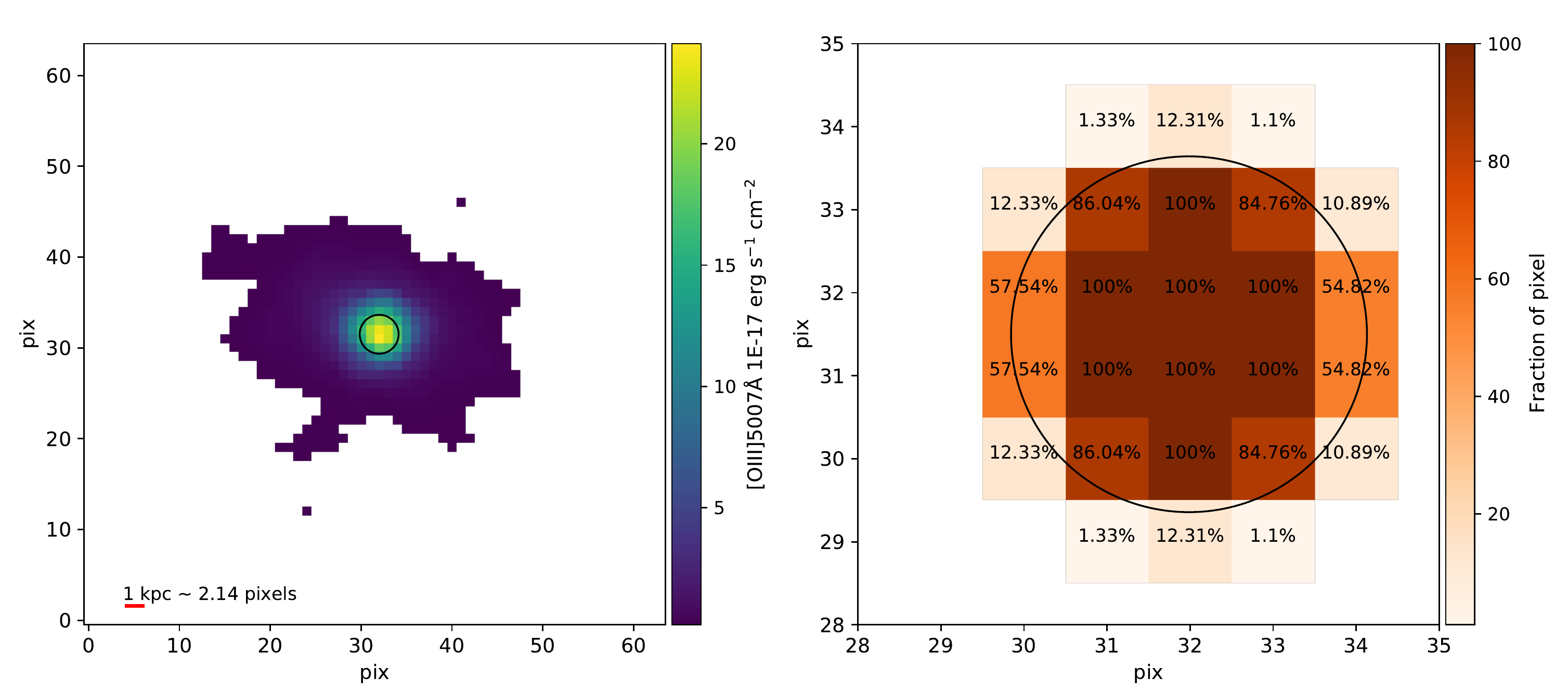}
  %[scale=1,bb=0 0 30 30]
  \caption{In the left plot we show (target with plate-ifu: 8725-9102) the flux of the [OIII]$\lambda$5007 emission line. Pixels with $S/N<3$ are masked. The black circle indicates the 2~kpc circular diameter aperture where we average the flux for the emission line ratios. In the right hand plot we show the zoom-in into the aperture region, showing the percentage of the pixel area captured by the aperture shown in black. We show the value in each pixel that corresponds to the weight of the pixel when we compute the average fluxes. }
         \label{fig_pix_weight}
\end{figure*}
%-----------------------------------------------------------------

\subsection{MaNGA DR17}

   Mapping Nearby Galaxies at Apache Point Observatory \citep[MaNGA,][]{bundy2015} is one of the surveys of the fourth generation of the Sloan Digital Sky Survey (SDSS-IV). It is an integral field unit (IFU) survey providing spatially resolved spectra \citep{Drory2015,Law2015} covering a spectral range from 3622 to 10354 \r{A} at a resolution R$\sim$2000 for each target, using the 2.5 Sloan Telescope \citep{Gunn2006}. The field of view of the IFU ranges from 12 to 32 arcseconds in diameter, ensuring that at least 80\% of the targets are covered out to 1.5 $R_{e}$ and 2.5 $R_{e}$, respectively. The 17th (and last) data release \citep{abdurrouf2021seventeenth} includes data for 10,010 unique galaxies at $0.01<z<0.15$ and stellar masses > 10$^{9}$ M$_{\odot}$.
   
   For MaNGA, a Data Reduction Pipeline (DRP) is provided in \cite{Law2016}, whose output is processed by the Data Analysis Pipeline \citep[DAP,][]{MangaDAP,Belfiore2019}. In the DRP, the raw data for each target is calibrated, sky subtracted, and stored in individual cube data and row-stacked spectra. The DAP then analyzes the latter to create cubes with the binned spectra together with models for the best fit spectra for different components (e.g. stellar continuum, emission lines). The DAP also provides maps of physical properties of the galaxies (e.g. sky coordinates, kinematics such as stellar and gas velocities, emission line fluxes, equivalent widths, and more). Through this paper, we use the emission line measurements (see Section \ref{sec3}) from the DAP for our analysis.
   
   A set of AGN catalogs (e.g. \citep{rembold2017,2018Wylezalek, sanchez2018, Comerford2020}) from previous SDSS-IV releases have provided AGN candidates since the early stages of MaNGA (e.g. MaNGA Product Launch 5, hereby MPL-5). These catalogs are described in the coming subsections and will be used for comparison purposes (see Section \ref{sec_comparisons}) to examine possible differences with our selection technique \ref{sec:optical_Select}. We will provide a detailed comparison of the ionized gas dynamics in AGN selected through different methods in an upcoming paper (Alb\'an et al., in prep.).

\subsection{MaNGA-MPL-5 AGN Catalog of S\'anchez et al.}
\label{mpl5sanchez}
   By June 2016, MaNGA had observed around 2700 targets \citep[MaNGA Product Launch 5 or MPL-5; the MPL5 sample is identical to the 14th data release of MaNGA,][]{SDSS-MaNGA-DR14}. For this sample, \cite{sanchez2018} found 98 AGN candidates. In this study, they used optical diagnostics (BPT diagrams), following the guidelines from \cite{kewley2006} on the three BPT classification diagrams ([N II]/H$\alpha$, [S II]/H$\alpha$ and [O I]/H$\alpha$). They focus on the emission-line fluxes inside a $3\arcsec$ aperture, derived using the PIPE3D \citep{sanchez2016} data analysis pipeline. In addition to the BPT diagnostics, they also include a cut in EW(H$\alpha$) of > 1.5\r{A} \citep[][]{fernandes2010}. A classification between type-I and type-II AGN \citep[see][for a detailed overview on AGN types]{Antonucci1993,Netzer2015} is also provided based on a multi-Gaussian emission-line fitting procedure in the spectral region containing H$\alpha$ and [N II]. They classify an AGN as type-I if the broad component satisfies a $S/N>5$ and $1000<\textnormal{FWHM}<10 000$ km/s. They identify 35 type-I AGN in their sample.

\subsection{MaNGA-MPL-5 AGN Catalog of Rembold et al.}
   An additional study by \cite{rembold2017} identifies 62 AGN candidates in the MaNGA MPL-5/DR14 galaxy sample. They follow a similar approach as in Section \ref{mpl5sanchez}, using optical-BPT diagnostics. However, the analysis is carried out using only one, namely the [N~II]-based BPT diagnostic. They use emission-line fluxes from a $3\arcsec\times3\arcsec$ aperture, but do not use the MaNGA data directly for their analysis. Instead, they utilise measurements from the SDSS-III single-fibre observations. The emission line fluxes and EW(H$\alpha$) were taken from the spectral analysis of \cite{thomas2013}, which requires an amplitude-over-noise (AoN) greater than two to calculate an emission line flux. They apply an EW(H$\alpha)>3\r{A}$ cut, as in \cite{fernandes2010}. The latter criterion is also used in our AGN selection.

\subsection{MaNGA-MPL-5 AGN Catalog of Wylezalek et al.}
\label{wylezalek_catalog}
    A different approach was used in \cite{2018Wylezalek}. Ionization radiation due to AGN can sometimes be found far away from the center of galaxies. Possible reasons for this effect are central obscuration, recent mergers, relic AGN \citep[e.g.,][]{Keel_2015}. Therefore, \cite{2018Wylezalek} develop a selection procedure based on spatially resolved BPT maps taking full advantage of IFU spectra and do not require AGN-like BPT diagnostics in the galaxy centers.
    
    In addition to the classical BPT line ratio diagnostics, they impose a suite of additional criteria to circumvent potential contamination of their sample through diffuse ionized gas, extraplanar gas, and photoionization by hot stars. They detect 303 AGN candidates in the same MaNGA Product Launch (MPL-5). They find 173 galaxies that would not have been selected as AGN candidates using the standard selection algorithms based on single-fibre spectral observations. 
    
\subsection{MaNGA-MPL8 multi-wavelength AGN catalog of Comerford et al.}
\label{Comerford_catalog}

\cite{Comerford2020} utilizes mid-IR, X-ray, and radio observations as well as broad emission lines in SDSS spectra to identify 406 AGN in the MaNGA MPL-8 (6261 galaxies) catalog. Their AGN catalog is thus independent of any BPT diagnostics. They find 67 AGN through mid-infrared selection criteria using data from the Wide-field Infrared Survey Explorer \citep[WISE]{Wright_2010}, 17 using hard X-ray selection criteria from Burst Alert Telescope \citep[BAT]{Barthelmy_2005} observations, and 325 AGN candidates through radio selection criteria using data from the NRAO Very Large Array Sky Survey \citep[NVSS]{Condon_1998} and the Faint Images of the Radio Sky at Twenty centimeters \citep[FIRST]{Becker_1995} survey. In this catalog, they also sub-divide radio-selected AGN into radio-loud and radio-quiet AGN. Additionally, they present 55 type-1 AGN candidates based on broad Balmer emission lines in SDSS spectra. Some galaxies are classified as AGN by several of these criteria (e.g., 11 galaxies were classified as AGN simultaneously by radio and mid-infrared selection criteria; see Table 2 in \citet{Comerford2020}).

%------------------------------------------------------------------
\section{AGN Selection}

   In this section, we present how IFU spectroscopy can be utilised in a variety of ways to classify MaNGA galaxies according to their BPT diagnostics. We extract the emission line properties of the targets in different aperture sizes and investigate how optical diagnostics vary and depend on the chosen aperture size. We finally present a suite of BPT-based MaNGA AGN catalogs with varying apertures for the full MaNGA sample (DR17).

  \begin{figure*}%[h!]
  \includegraphics[width=\textwidth]{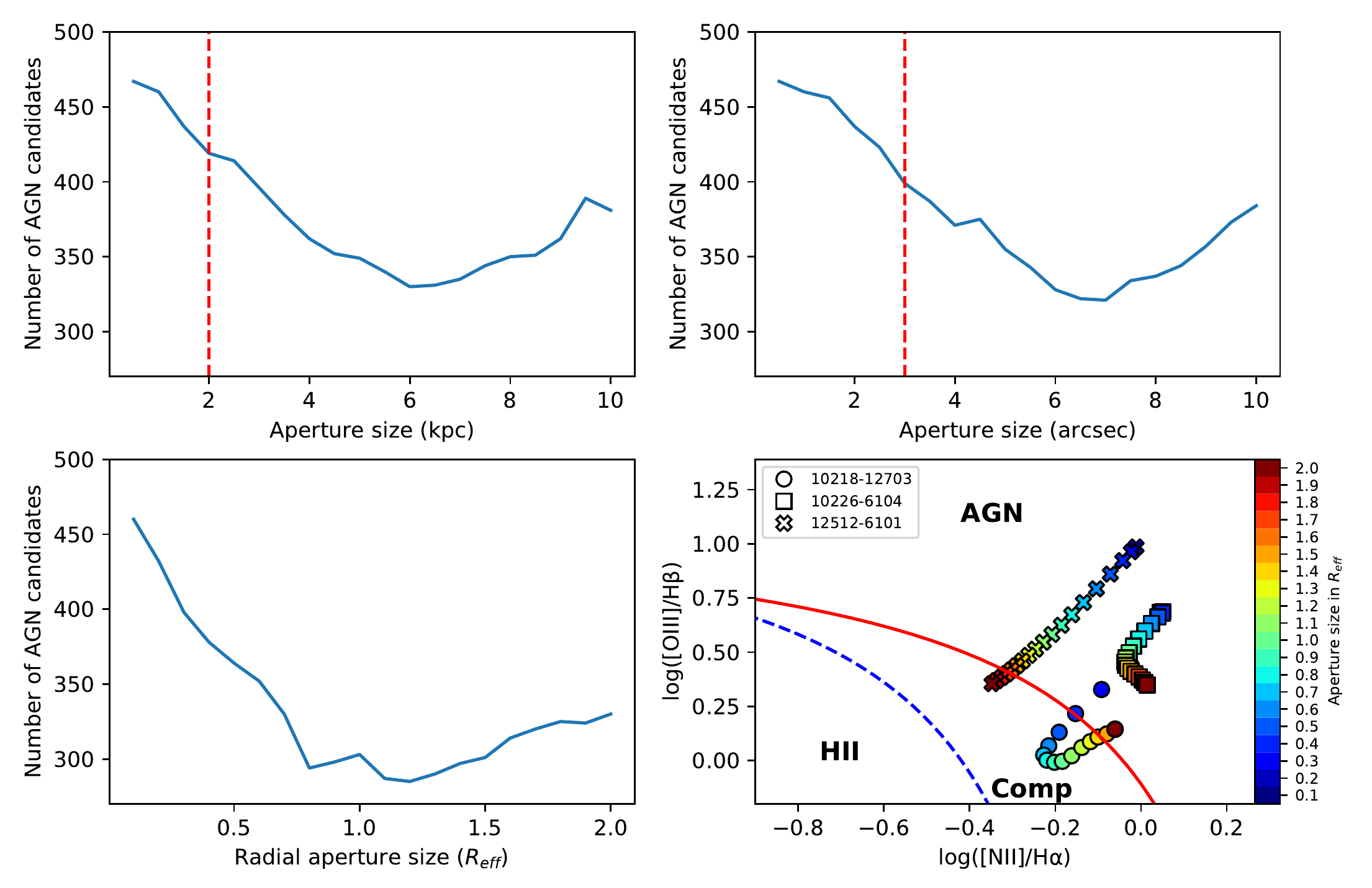}
  \caption{The y-axis in the top plots and the bottom left plot shows the number of galaxies classified as AGN (solid blue line) selected based on different aperture sizes and using different units. The selection is done following the algorithm described in Section \ref{sec:optical_Select}, excluding the [O~I]/H$\alpha$ BPT diagram. In the top left plot, the kpc aperture increases in steps of 0.5~kpc (in diameter). In the top right plot, the arcsecond aperture increases in steps of 0.5$\arcsec$ (in diameter). In the bottom left plot, the effective radius aperture increases in steps of 0.1~R$_{eff}$ (in radius). The red (vertical-dashed) line in the top panels corresponds to the apertures used for the comparisons in Section \ref{sec_comparisons}. In the bottom right plot, we show how the flux ratios in one BPT diagram change as the aperture does. This is done for three individual galaxies illustrated with different symbols (check the legend for their corresponding MaNGA-IDs). Each symbol is filled with a specific color that corresponds to the size of its aperture. In the bottom right plot, we also include the empirical division lines that will give each target a specific classification (e.g., AGN-like galaxy, Composite object, or HII-star forming galaxies): red lines correspond to \cite{kewley2001}, and the blue dashed line on the left plot corresponds to \cite{Kauffmann2003}.}
  \label{how_much_AGN}
         
\end{figure*}

\label{sec3}
%------------------------------------------------------------------

\subsection{Sample Definition}
\label{sample_Select}

   To obtain the emission line fluxes, as required by a BPT classification, we extract the emission line maps for [O~III]$\lambda$5008, [N~II]$\lambda$6584, [S~II]$\lambda\lambda$6717,6731 (hereafter, we will refer to the sum of [S~II]$\lambda\lambda$6717,6731 simply as [S~II]), [O~I]$\lambda$6300 and the hydrogen Balmer lines H$\alpha$ and H$\beta$ from the DAP. We use their non-parametric emission line measurements. The latter are derived using bands of 20 \r{A} centered on each emission line, and a narrower passband is used for H$\alpha$ and [N~II], due to their small separation \citep[see][]{MangaDAP}. No special treatment is done for possible complexities in emission lines such as galactic outflows \citep[e.g.,][]{Wylezalek2020} or emission from the broad line region \citep[BLR,][]{Peterson_2006,Netzer2015}. We use the MANGA\_DAPPIXMASK\footnote{You can find information on how to use these masks in the following link: https://www.sdss.org/dr17/algorithms/bitmasks/} bitmap to mask pixels flagged as DONOTUSE or UNRELIABLE to exclude biased and/or unreliable measurements.
   
   Furthermore, in each map, we mask the pixels where the emission line flux does not satisfy a signal-to-noise ($S/N$) greater than three (the $S/N$ is also extracted from the DAP). Emission lines with a $S/N<3$ lead to unreliable flux measurements and thus unreliable emission line ratios which would increase the bias in the classification \citep[e.g.,][]{Brinchmann2004}. We measure the pixel-weighted flux before computing the emission-line ratios. When a pixel (at a certain position) in a map of a specific parameter (e.g., the [N~II] flux) has a $S/N<3$, we do not use that pixel for computing the emission line flux of that aperture (see Section \ref{sec:optical_Select}). However, if that pixel (same position as before) has enough signal-to-noise ($S/N>3$) in another map (e.g., the [S~II] flux) of the same galaxy, it is included for computing that emission line flux.

  This $S/N$ cut inevitably decreases the size of our galaxy sample. For example, when using a 2~kpc-sized aperture (see Section \ref{aperture_based}), in $\sim$25\% of the MaNGA galaxies no pixels are left available for measuring emission line fluxes due to the signal-to-noise cut. Therefore, our sample is biased towards line emitting galaxies, which are preferentially gas-rich, star-forming galaxies. We explore this bias further in Section \ref{sec_em_line} and in Section \ref{sec_no_em_line}.

  \subsection{Optical Classification}
  \label{sec:optical_Select}
   Due to the IFU nature of the MaNGA observations, we can perform an aperture-dependent optical classification. We perform weighted averages in a way that if a pixel partially contributes to a circular (or squared) aperture, we weigh the pixel according to the fraction of its enclosed area. The latter is measured as follows:
   \begin{equation}
   \label{eq_weighted_mean}
       {\displaystyle {\bar {x}}={\frac {\sum \limits _{i=1}^{N}x_{i}w_{i}}{\sum \limits _{i=1}^{N}w_{i}}},}
   \end{equation}
   where x$_{i}$ is the total contribution of the pixel and w$_{i}$ is the weight of the pixel measured as the fraction of the enclosed area by the corresponding aperture for the $N$ enclosed pixels. We construct a squared grid to represent the area and positions of each pixel. Each pixel is placed in the center of the squares of the grid, evenly distributed according to the average distance between the original coordinates (as in the coordinate map known as SPX\_SKYCOO, from the DAP maps). 
   
   We compare the coordinate of the center of each square (from the grid) to the original center (from the DAP map SPX\_SKYCOO) of the pixel. The typical offsets between the coordinates of our synthetic grid and the ones from the DAP are of the order of 10$^{-8}$$\arcsec$, and the distances from each pixel are of the order of 10$^{-2}$$\arcsec$, demonstrating that the deviations from the original position of the pixels are negligible. Thus, we do not take into account this effect when calculating average emission line fluxes inside the respective apertures. An example of this interpolation is shown in Figure \ref{fig_pix_weight}, where in the left plot we display an example of the [O~III]$\lambda$5008 emission-line flux map (masked as described in Section \ref{sample_Select}) with the aperture shown with a black circle (whose size corresponds to 2~kpc). In the right plot we show a zoom-in version of this map using a different color map based on the fraction of the pixel enclosed by the aperture (again shown in black); note that pixels whose area enclosed by the aperture is zero are masked as NaN values and are excluded from the plot. To create a specific aperture, the digital size of a pixel is converted to physical quantities using data from the DAP (e.g., effective radius, pixel per arcsecond and kiloparsec steps) and the DRP (for individual redshifts using the NSA redshift measurements).
  
   We measure average emission line fluxes in a set of apertures following the quality criteria outlined in Section 3.1. We then compute the following averaged flux ratios: [O~III]/H$\beta$, [N~II]/H$\alpha$, [S~II]/H$\alpha$ and [O~I]/H$\alpha$ (also with a $S/N>3$ cut).  For a specific aperture (e.g. 2~kpc or $3\arcsec\times 3\arcsec$), we perform two separate classifications \citep{kewley2006,Kauffmann2003,kewley2001}: one using all three BPT diagrams and one excluding the [O~I]6300/H$\alpha$ BPT diagram. The [O~I]6300 emission line is relatively weaker compared to the other lines in the BPT diagrams and therefore exhibits lower S/N leading. This would therefore lead to the exclusion of many more galaxies from the analysis (see Table \ref{table1} and Table \ref{table2}). If a galaxy is classified as one type in one diagram and as a different type in the remaining diagrams (whether we are using two or three of the diagrams), it will be referred to as Ambiguous.
Additionally, we also apply the diagnostic criteria outlined by \citet{fernandes2010} which allows the differentiation between two very distinct classes of galaxies that overlap in the LINER region of the BPT diagrams, namely galaxies hosting a weak AGN and "retired galaxies". Retired galaxies have stopped forming stars and are ionized by their hot evolved low-mass stars, i.e. post-AGB stars. This differentiation can be achieved by using an equivalent width (EW) cut of EW($H\alpha)>3$~\r{AA}. We use the EW(H$\alpha$) from the DAP and also measure its average (see equation \ref{eq_weighted_mean}) in the various apertures. We apply this EW($H\alpha$) cut to all galaxies above star-forming demarcation lines \citep{kewley2006}. This step concludes the classification procedure. As mentioned in Section 3.1., not all galaxies could be classified due to the quality cuts. We will refer to these low $S/N$ galaxies as the unclassified ones (see Section \ref{sec_no_em_line}).
   
   We point out that we have not taken into account the inclination angles of the individual galaxies, meaning that all the circular or squared apertures are performed on the line-of-sight projected galaxy. 

   \begin{figure}[t]
\centering
\includegraphics[width=\hsize]{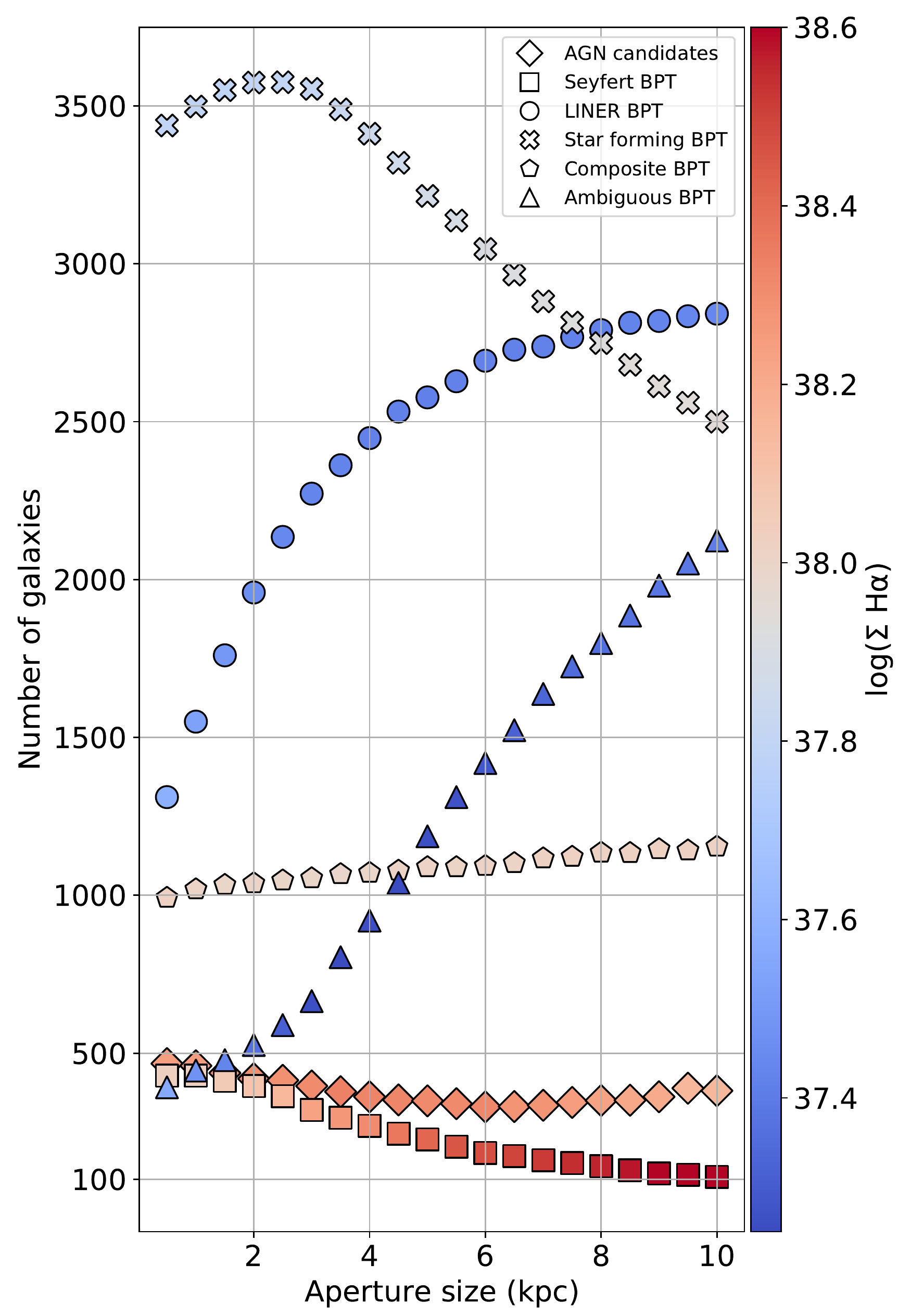}
  \caption{Number of galaxies (y-axis) identified by a specific BPT classification (see the legend on the top right of the plot) using different selection aperture sizes (in kpc, x-axis). Each value is colored by the average H$\alpha$ surface brightness (color-bar) within a 2~kpc aperture. This classification is done excluding the [O~I] BPT diagram.}
     \label{alltypes_with_sfb}
\end{figure}

   \begin{figure}[h!]
\centering
\includegraphics[width=\hsize]{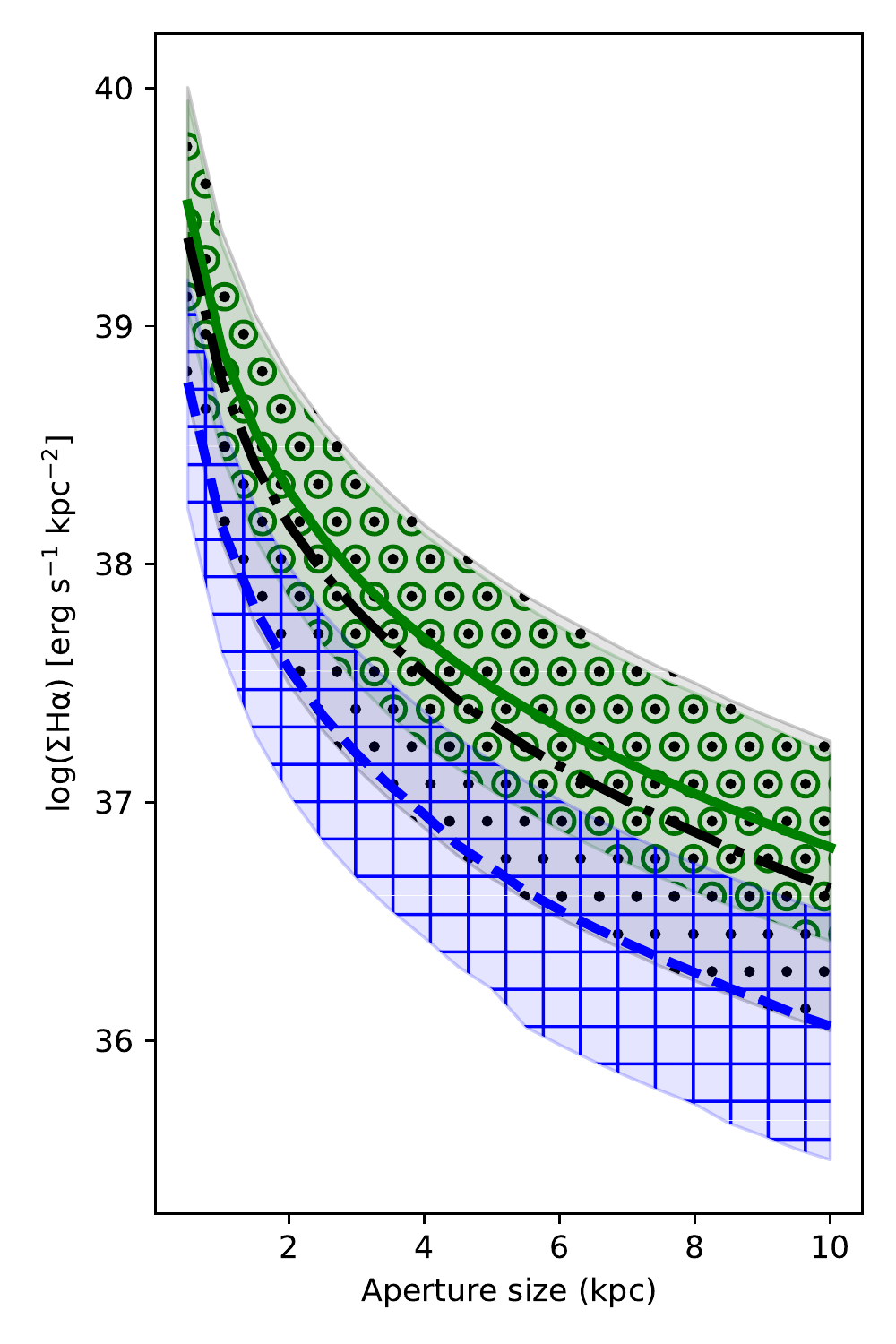}
  \caption{Average radial profiles of the H$\alpha$ surface brightness stacked according to specific AGN candidates. The shaded regions correspond to the 16th and 84th percentiles of each stacked radial profile. The solid green line corresponds to the AGN selected using a 2~kpc circular aperture, and hollow circles hatch its shaded area. The black dash-dot line corresponds to the AGN selected by an aperture of 10~kpc, and dots hatch its shaded area. Finally, the blue dashed line corresponds to the AGN selected by the 10~kpc aperture but excludes the AGN selected by the 2~kpc aperture. The last line has a shaded area hatched by hollow squares.}%
     \label{radial_surf_dig_effect}
\end{figure}

\subsection{Aperture effects on BPT AGN Selection}
\label{aperture_based}

   In this Section, we explore the behavior of an aperture-dependent classification. To do so, we investigate the BPT classification as described in Section \ref{sec:optical_Select} in 60 different apertures of different sizes, defined as follows:
    
   \begin{itemize}
      \item 20 circular-shaped apertures with sizes ranging from 0.5~kpc to 10~kpc in diameter in steps of 0.5~kpc .
       \item 20 squared-shaped aperture with sizes ranging from 0.5$\arcsec$ to 10$\arcsec$ on the side in steps of 0.5~$\arcsec$.
      \item 20 circular-shaped apertures with sizes ranging from 0.1~R$_{eff}$ to 2~R$_{eff}$ in diameter in steps of 0.1~R$_{eff}$.
   \end{itemize}

    In Figure \ref{how_much_AGN} (bottom right plot), we show an example of the behavior of the emission line ratios from three different targets when using 20 circular-shaped apertures. This reveals the dependence of the aperture on our classification. In Figure \ref{how_much_AGN} (see the two upper plots and the one on the bottom left) we show how the number of galaxies classified as AGN changes as a function of an increasing kpc-, arcsecond- and effective radius (R$_{eff}$)-based aperture, respectively. For simplicity, we use the term `AGN' to refer to `galaxies classified as AGN' hereafter. 

    As expected, in the smallest apertures, independent of the specific unit, the number of galaxies classified as AGN is similar. The number of AGN in R$_{eff}$-based apertures, though, drops steeply (compared to the other apertures) and reaches its minimum at $\sim$0.7-1~R$_{eff}$. This happens because the step that we use based on effective radius aperture reaches the outerskirts of the galaxies faster, leading to fewer galaxies classified as AGN. In the $\sim$7500 galaxies that we were able to classify (see Table \ref{table1}), the average R$_{eff}$ corresponds to $\sim$4.63~kpc, meaning that an aperture of 0.7~R$_{eff}$ will have an average size of $\sim$6.5~kpc in diameter. 
    
   We also observe an initial decrease in the total number of classified AGN in the kpc- and arcsec-based apertures. We then observe an increase in classified AGN at larger aperture sizes in all panels. The increasing number of AGN at larger aperture sizes is most notably seen in the kpc- and arcsec-based apertures (see the top panels of Figure \ref{how_much_AGN}) beyond $\sim$6~kpc and $\sim$7$''$, respectively (see Section \ref{sec_digs}). In the R$_{eff}$-based apertures, the number of AGN increases beyond $\sim$1.2~R$_{eff}$.
   
   For all the targets in MaNGA DR17, we provide FITS tables containing all of our relevant measurements (emission line ratios and H$\alpha$ EWs, if available) in all here investigated apertures. We report the BPT class of each galaxy (before applying the EW cut and excluding the [O~I]6300/H$\alpha$ diagram) and the resulting AGN catalog for each specific aperture. We provide these catalogs as part of the supplementary material. We show an example of one of our catalogs in Table \ref{table:catalog}, specifically for the 2~kpc-based classification. Note that (as discussed in Section \ref{sec:optical_Select}) for a galaxy to be classified as an AGN, our procedure not only requires the emission-line ratios to be located above the star-forming demarcation lines \citep[see][]{kewley2006} but also to have an EW(H$\alpha$) greater than 3 (whether they are selected as Seyfert or LINER). For example, in row number 8528 the target is classified as Seyfert initially, but because of a low EW(H$\alpha$) this target fails to meet our final AGN criteria. The same is true for the LINER galaxy in row number 8529. Due to our $S/N$ criteria, some values are stored as NaN, and thus we do not assign any classification (e.g. 8531, note that these targets do not even enter the Ambiguous BPT classification). In row number 8535, a Seyfert did meet the EW(H$\alpha$) cut and is selected as an AGN candidate. In that case, the [O~I]/H$\alpha$ column is NaN, but this does not impact the classification as [O~I]/H$\alpha$ is not used here. The classifications and measurements in these catalogs can be used according to the science goals. The most straightforward advantage is being able to choose parameters and classifications from a specific aperture. For example, one can alternatively decide to relax on the EW(H$\alpha$) criterion when selecting AGN (if needed), and select AGN from the Seyfert and LINER population with EW(H$\alpha$)$>1.5$~\r{AA}. Therefore, it is possible to easily extract modified classifications from these catalogs.
   %\begin{sidewaystable}
\begin{table*}
\caption{BPT classification (without using [O~I]/H$\alpha$), AGN selection, emission line-ratios, and equivalent widths values for the MaNGA galaxies. This is one out of 60 tables, each corresponding to a different aperture (see Section \ref{aperture_based}). We show our selection considering a 2~kpc aperture. From the first to the last column, the table is shown as it is stored in the supplementary data. The first column reports the ID of the row. The second column reports the Plate and IFU of the corresponding target. In the third column, a boolean array is stored to flag a galaxy as an AGN if the value is 1 and 0 if it is not an AGN \textbf{(i.e., Seyferts and LINERs with EW(H$\alpha)>3$~\r{AA})}. From the fourth to the eighth column, the BPT class is shown. We abbreviate Seyferts, LINERs, Star-forming, Composite and ambiguous classification as SY, LI, SF, CM, and AM, respectively (however, in the fits table, they are tagged as their full name). From the ninth to the twelfth column, the emission line flux ratios are given as logarithmic bases of 10. Lastly, the last column stores the information of the equivalent width in \r{A}. If a value is tagged as $nan$, the emission line did not pass our S/N criteria (see Section \ref{sec:optical_Select}.)}
\label{table:catalog}      
\centering          
\begin{tabular}{c c c c c c c c c c c c c }     % 7 columns 
\hline%\hline       
                      % To combine 4 columns into a single one 
ROW & MANGAID & AGN & SY & LI & SF & CM & AM & [O~III]/H$\beta$ & [N~II]/H$\alpha$ & [S~II]/H$\alpha$ & [O~I]/H$\alpha$ & EW\_H$\alpha$ \\ 
 & Plate-IFU &  \multicolumn{6}{c}{Boolean array: $0=$~False~$\hdots$~$1=$~True} & log10 & log10 & log10 & log10 & \r{A} \\ 
\hline                    
0 & 10001-12701 & 0 & 0 & 0 & 1 & 0 & 0 & -0.03 & -0.56 & -0.37 & -1.32 & 22.25 \\
1 & 10001-12702 & 0 & 0 & 0 & 0 & 1 & 0 & -0.18 & -0.3 & -0.23 & -0.63 & 5.03 \\
\vdots &\vdots &\vdots &\vdots &\vdots &\vdots &\vdots &\vdots &\vdots &\vdots &\vdots &\vdots & \vdots  \\
8526 & 9040-3704 & 0 & 0 & 0 & 0 & 0 & 1 & 0.12 & -0.22 & -0.28 & -1.03 & 7.35 \\
8527 & 9040-6101 & 0 & 1 & 0 & 0 & 0 & 0 & 0.65 & 0.17 & -0.07 & nan & 1.07 \\
8528 & 9040-6102 & 0 & 1 & 0 & 0 & 0 & 0 & 0.36 & -0.05 & -0.28 & nan & 2.22 \\
8529 & 9040-6103 & 0 & 0 & 1 & 0 & 0 & 0 & 0.28 & 0.02 & 0.2 & 0.08 & 0.3 \\
8530 & 9040-6104 & 0 & 0 & 0 & 1 & 0 & 0 & -0.48 & -0.34 & -0.69 & -1.51 & 9.8 \\
8531 & 9040-9101 & 0 & 0 & 0 & 0 & 0 & 0 & nan & nan & nan & nan & nan \\
8532 & 9040-9102 & 1 & 0 & 1 & 0 & 0 & 0 & 0.09 & 0.09 & -0.17 & -0.78 & 6.17 \\
8533 & 9041-12701 & 0 & 0 & 0 & 1 & 0 & 0 & -0.49 & -0.39 & -0.59 & -1.61 & 14.0 \\
8534 & 9041-12702 & 0 & 0 & 0 & 0 & 1 & 0 & -0.31 & -0.32 & -0.52 & -1.33 & 6.3 \\
8535 & 9041-12703 & 1 & 1 & 0 & 0 & 0 & 0 & 0.49 & -0.07 & -0.27 & nan & 3.47 \\
\vdots &\vdots &\vdots &\vdots &\vdots &\vdots &\vdots &\vdots &\vdots &\vdots &\vdots &\vdots & \vdots  \\
10241 & 9894-9102 & 0 & 0 & 0 & 1 & 0 & 0 & 0.51 & -1.32 & -0.71 & -1.54 & 114.6 \\

\hline                  
\end{tabular}
%\end{sidewaystable}
\end{table*}

   \begin{figure}[t!]
\centering
\includegraphics[width=\hsize]{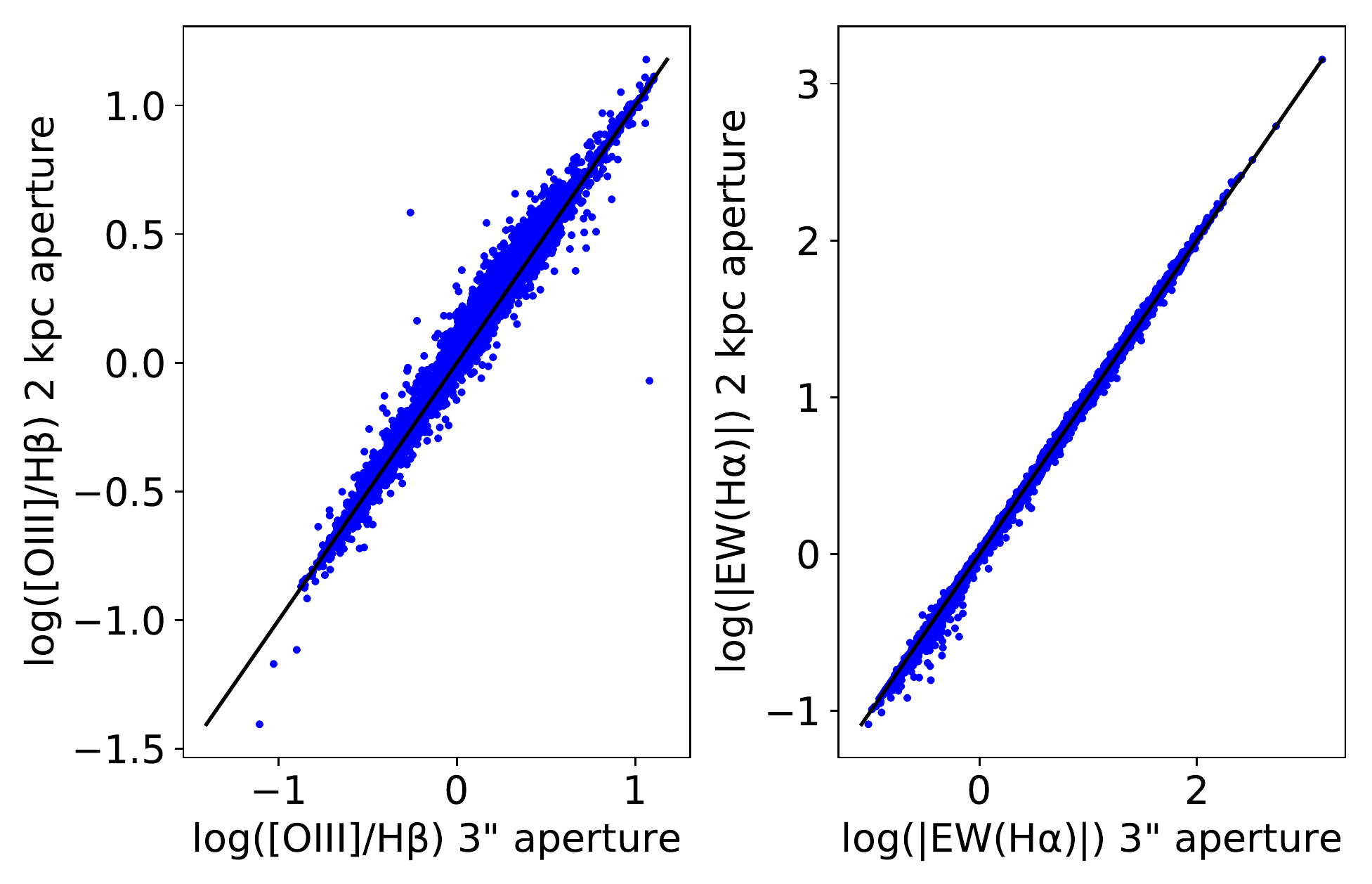}
  \caption{Filled blue circles correspond to a comparison between measurements of EW(H$\alpha$) (right-hand plot) and [OIII]/H$\beta$ ratio (left-hand plot) from our 2~kpc aperture (y-axis) and $3\arcsec\times 3\arcsec$ aperture (x-axis). The measurements shown here correspond to the full MaNGA sample. The solid black line in both plots shows the one-to-one ratio that the values should follow if they had the same value.}
     \label{comparison_y_own}
\end{figure}

\subsection{Aperture effects on BPT type classification}
\label{agn_types}

  In Figure \ref{alltypes_with_sfb} we show how the classification in individual BPT sub-types (i.e., star-forming, Seyfert, LINER, Composites, Ambiguous) changes as a function of aperture size. For this analysis, we choose the BPT classification only using the [N~II] and [S~II] BPT diagrams along derived from a kpc-based aperture (the color coding corresponds to the H$\alpha$ surface brightness, which we will discuss in detail in Section \ref{sec_digs}). Specifically, at an aperture size of $\sim$ 2~kpc, the number of star-forming galaxies reaches its maximum value, while the number of Ambiguous objects and the number of Seyfert galaxies decreases beyond this aperture. At very large aperture sizes, we see a drastic decrease in the galaxies classified as star-forming as the number of galaxies classified as LINER and Ambiguous galaxies increases. This effect is explored in Section \ref{sec_digs} and in Section \ref{sec_em_line}. 
  
\section{Analysis}
\label{sec_results}

\subsection{The role of diffuse ionized gas}
\label{sec_digs}

   Early studies of radio observations and optical emission lines have provided evidence for the existence of diffuse ionized components outside bright H~II regions and in kpc-scale layers extending above 1-2~kpc of the Galactic Plane as well as in nearby galaxies \citep{Hoyle1963,Reynolds1984,Rand1990}. The diffuse ionized gas \citep[DIG; see][for a review]{Haffner2009} is an important element of the ISM \citep[and currently an active field of study; e.g.,][]{Belfiore2016,Zhang2016,Jones2017,2018Wylezalek,Asari2019,Krishnarao2020,Mannucci2021}, corresponding to a widespread (low density $\sim$ 0.1~cm$^{-3}$), warm gas (10~000 K). \cite{Zhang2016} has shown that DIG has typically low H$\alpha$ surface brightness and can significantly affect emission line ratios and thus the typical optical diagnostics, suggesting that choosing galaxies / spaxels with $\Sigma_{H\alpha}$>10$^{39}$~erg~s$^{-1}$~kpc$^{-2}$ will result in more reliable H~${II}$ dominated regions. DIG not only can drive true star-forming galaxies to the Ambiguous and LINER-like BPT classification regime but also can imitate AGN-like emission line ratios.

   Although integral field unit surveys provide an excellent observational tool for studying the spatially resolved properties of galaxies, individual MaNGA spaxels resolve properties down to the kpc scales. Thus, it is inevitable that multiple ionizing sources contribute to individual spaxels, which include contamination from DIG-dominated regions. In Figure \ref{how_much_AGN} we see that apertures exceeding 6~kpc, 7$''$, and 1.2~$R_{eff}$ start to detect `AGN candidates' at larger radii. As discussed in Section \ref{arc_kpc_catalog}, we expect reliable AGN signatures from the inner kpc regions of true AGN candidates. Therefore, we explore the potential role of the DIG in affecting the increasing AGN detection rate at larger apertures.
   
      To do so, we measure the cumulative\footnote{We checked the stacking of cumulative and non-cumulative radial profiles of H$\alpha$ surface brightness (for AGN candidates). The negligible differences do not impact the discussion in this section. We focus on the cumulative contribution of this parameter motivated by taking into account the effects of an increasing aperture-based selection.} radial profiles of the H$\alpha$ surface brightness for all MaNGA targets following the procedure in Section \ref{sec:optical_Select}. We stack the cumulative radial profiles of the AGN-selected galaxies from a 2~kpc (diameter) aperture (419 galaxies) and a 10~kpc aperture (381 galaxies). Additionally, we compute the stacked radial profile using the AGN candidates that were selected by the 10~kpc aperture, excluding the AGN that were selected in both the 2~kpc and 10~kpc aperture (158 galaxies). The latter was done only using the [N~II] and [S~II] BPT diagrams. This is shown in Figure \ref{radial_surf_dig_effect}. All the AGN selected galaxies from a 2~kpc (diameter) aperture (green solid line) show a high central H$\alpha$ surface brightness ($\Sigma_{H\alpha}\sim$~10$^{39.5}$~erg~s$^{-1}$~kpc$^{-2}$) and a higher cumulative radial profile when compared to the AGN selected only based on the 10~kpc aperture (i.e., excluding the ones selected from the 2~kpc aperture). These 10~kpc AGN candidates display a lower central surface brightness ($\Sigma_{H\alpha}$<10$^{39}$~erg~s$^{-1}$~kpc$^{-2}$). Both populations (2~kpc and 10~kpc selected) show significantly different levels and distributions of H$\alpha$ surface brightness at every radius, with almost a dex of difference.
   
   We relate this systematic decrease of the H$\alpha$ surface brightness to be consistent with an increasing contribution from DIG regions. This explains the increase of galaxies classified as AGN in larger apertures observed in Figure \ref{how_much_AGN}, which suggests that an increasing aperture would lead to the inclusion of a DIG-biased population of AGN candidates. We further study this in Figure \ref{alltypes_with_sfb}, where we show the number of AGN candidates with a specific classification (each represented with a different shape) when using different selection apertures. The color map in this Figure shows the average H$\alpha$ surface brightness measured in a 2~kpc aperture for each data point. Here, the number of targets classified as LINER (circles) and Ambiguous (triangles) galaxies increases at larger apertures while the average H$\alpha$ surface brightness (measured at a 2~kpc aperture) decreases and is generally at a low level.  This is consistent with the results from \citet{Belfiore2016}, where Low ionization emission-line regions (LIER) dominate around $\Sigma_{H\alpha}\sim$~10$^{38}$ \citep[see Figure 7 in][]{Belfiore2016}, suggesting that these objects may be strongly contaminated by DIG regions.
   
   Objects classified as Composite, AGN, Seyferts or star forming galaxies have higher central H$\alpha$ surface brightness at all apertures ($\Sigma_{H\alpha}\gtrsim 10^{38}$). Galaxies classified as Composites have roughly a constant central H$\alpha$ surface brightness at all apertures. We observe that objects that are still classified as Seyfert galaxies when selected from very large kpc apertures are the ones who show the highest central H$\alpha$ surface brightness. This is very likely due to the fact that the contamination from DIG at larger radii must be compensated by higher central H$\alpha$ surface brightness at the center for a galaxy to still make it into the Seyfert regime of the BPT classification.

\begin{figure}[t]
\includegraphics[width=\hsize]{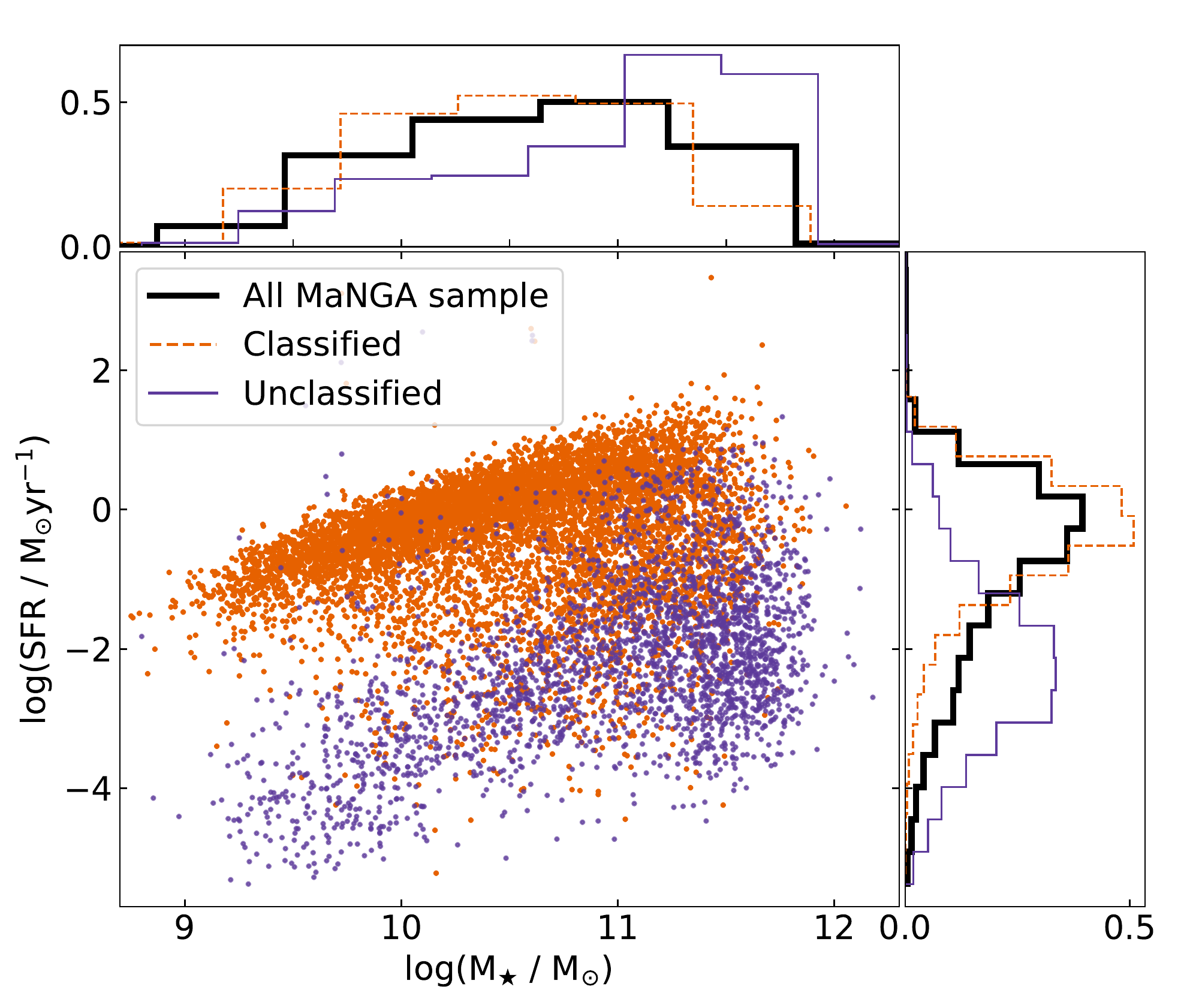}
  \caption{For all MaNGA galaxies, we show the stellar mass versus star formation rate (derived from H$\alpha$), extracted from the Pipe3D \citep{sanchez2016} catalog. We color each target in orange or purple respectively on the scatterplot to determine whether or not that specific galaxy was considered in the BPT classification scheme (meaning that it met our $S/N$ criteria specified in Section \ref{sample_Select}). This plot is based on the classification using a 2~kpc aperture without considering the [O~I] BPT diagram. The histograms show the distributions of stellar mass (top) and star formation rate (right). Distributions for the BPT classified and unclassified galaxies correspond to the dashed and solid histograms, respectively. The distribution of the whole MaNGA sample is shown in the black bold histogram. The distributions of the histograms are normalized, and the y-axis in the histograms corresponds to the normalized counts.}
     \label{fig_gals_lost}
\end{figure}

\begin{figure}[t]
\includegraphics[width=\hsize]{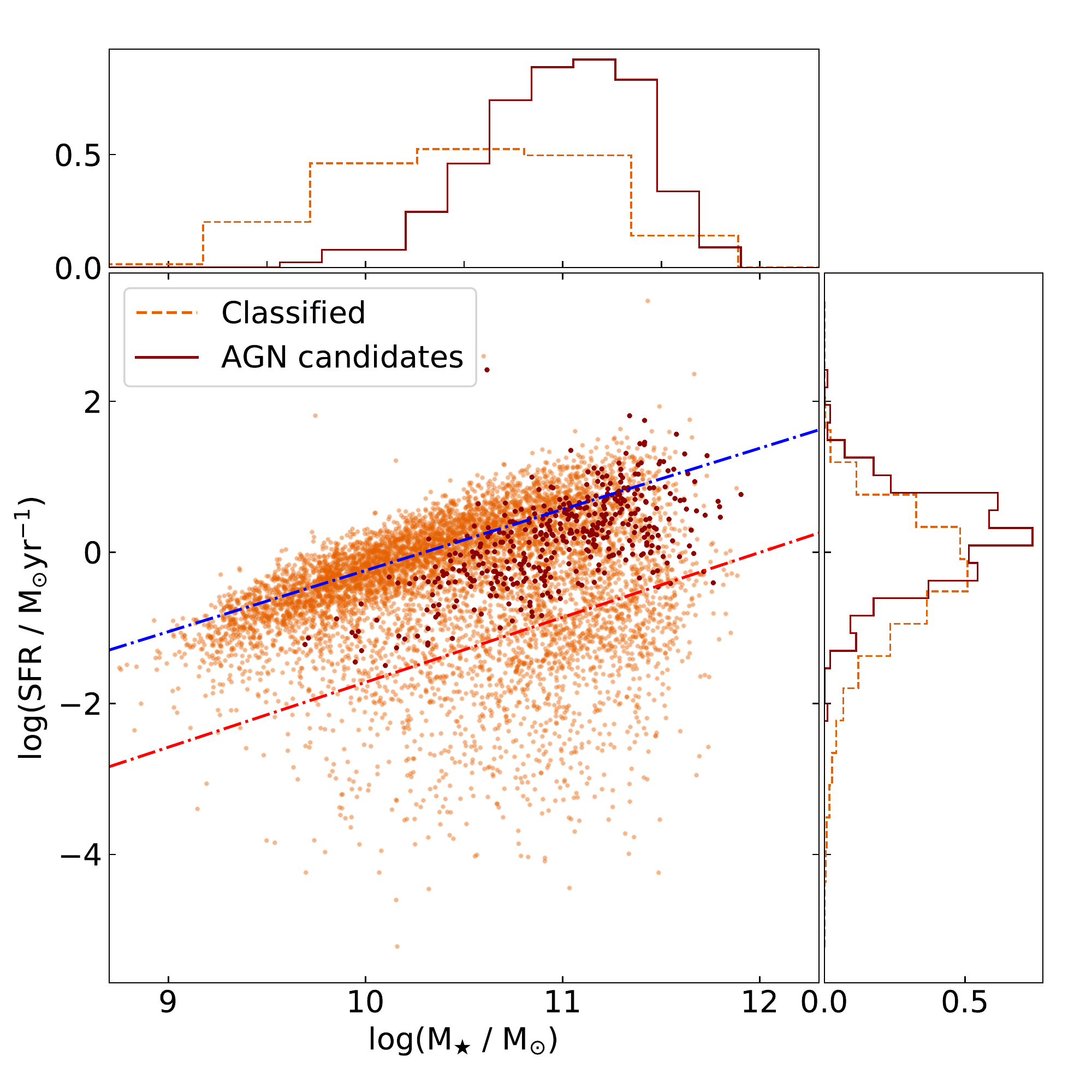}
  \caption{Similar to Figure \ref{fig_gals_lost}, we focus here on the BPT classified sample (orange), highlighting the AGN candidates (dark red) selected from a 2~kpc aperture (see Section \ref{sec:optical_Select}). We show the main sequence for star-forming galaxies (SFMS) and the Retired Galaxies Sequence (RGS) from \citet{Cano_D_az_2016} using blue and red dashed lines, respectively.}
     \label{fig_AGNs_MS}
\end{figure}

\subsection{Classifications based on a 3$\arcsec$ and 2~kpc aperture}
\label{arc_kpc_catalog}

   Many current and early surveys have studied the general properties of galaxies based on a single fibre 2$\arcsec$ \citep[e.g.,][]{Dawson2013} or 3$\arcsec$ apertures \citep[e.g.,][]{Gunn2006}. Consequently, AGN identifications have been frequently done, especially for galaxies observed within SDSS surveys, based on measurements in these apertures. For MaNGA, for example, key AGN work has been done by classifying AGN using optical diagnostics on a 3$\arcsec$ aperture \citep{rembold2017,sanchez2018}. We, therefore, select the same aperture (3$\arcsec$) to study the impact of low redshift optically-selected AGN candidates when compared to a kpc-based aperture catalog. Our kpc-based aperture is used with the intention of reducing aperture effects (given by the redshift range of MaNGA galaxies) on selected AGN candidates and to focus on the compact AGN region.
   
   AGN can ionize the gas in host galaxies up to kpc scales. This is commonly referred to as the narrow line region \citep[NLR, see][]{Netzer2015}. The typical sizes of the NLR have been studied over the past decades \citep[e.g.,][]{Bennert2006,Chen2011,Netzer2015,Padovani2017}, with a consensus of a lower limit between hundreds of pc up to $\sim$1~kpc of radius. Therefore, the ionizing source of some AGN can be easily diluted even at small apertures.
   
   We note that the smaller the aperture, the more galaxies are classified as AGN (see Figure \ref{how_much_AGN}). Specifically, we find that 59 AGN candidates from the 0.5~kpc catalog are not classified as AGN in the 2~kpc catalog. The median redshift of these 59 galaxies is ~0.032, while the median redshift of the AGN population at 2~kpc is ~0.045. This corresponds to a median resolution of ~1.05~kpc and ~1.58~kpc, respectively\footnote{We can estimate the median spatial resolution of these AGN-candidate population using the spatial element relation $\delta\approx35z$~kpc \citep[see][]{Bundy_2014}.}. Indeed, this suggests that the NLR sizes of these 59 AGN candidates (if proven to be true AGN at all) are small enough to not be resolved by the MaNGA survey and therefore get diluted when using a 2~kpc aperture. If we perform the same comparison between the 2~kpc and 5~kpc catalog, we find that 122 galaxies are not classified as AGN in the 5~kpc catalog. The median resolution of these 122 galaxies is 1.12~kpc and the median resolution of the AGN population in the 5~kpc catalog is 1.60~kpc. When we carry out such comparisons now using the arcsecond or effective radius-based apertures, we find that the galaxies that are not classified as AGN in larger apertures still have a similar median kpc resolution compared to the ones at smaller apertures. We interpret this as a clear suggestion that the kpc-based aperture is more consistent with the physical properties of AGN.

   Furthermore, MaNGA samples can achieve a median spatial resolution of about 1.37 kpc \citep{Wake2017}. This spatial resolution and the suggested physical sizes of the NLR motivate us to use an aperture of 2~kpc in diameter. To carry out a comparison between catalogs based on different apertures (see Section \ref{sec_comparisons}), we choose the following four catalogs:
   
   \begin{itemize}
      \item circular-shaped aperture with a 2~kpc diameter using full BPT diagnostics ([NII]/H$\alpha$, [SII]/H$\alpha$, [OI]/H$\alpha$).
      \item circular-shaped aperture with a 2~kpc diameter excluding the [OI]/H$\alpha$ BPT diagnostic.
      \item square-shaped aperture with a 3$\arcsec$ size on the side using full BPT diagrams ([NII]/H$\alpha$, [SII]/H$\alpha$, [OI]/H$\alpha$).
      \item square-shaped aperture with a 3$\arcsec$ size excluding the [OI]/H$\alpha$ BPT diagnostic.
   \end{itemize}

   We compute each AGN catalog following the procedure described in Section \ref{sec:optical_Select} including the EW(H$\alpha$) cut. The full classification and the number of galaxies corresponding to each class (Star-forming, Seyfert, LINER, Composite and Ambiguous, and the final AGN candidates) for these apertures is shown in Table \ref{table1} and \ref{table2}, as well as the $Total$ number of galaxies that were used in that BPT classification. In these tables, we also report the number of galaxies that the different selections have in common. 
   
\subsection{Hosts of emission line galaxies}
\label{sec_em_line}

   The number of successfully BPT-classified galaxies is an increasing number of decreasing $S/N$ cuts. The BPT classification becomes more biased towards less massive galaxies \citep{Brinchmann2004} as a function of increasing $S/N$ cut (see below). To present the impact of using our $S/N$ limit ($S/N>3$), in Figure \ref{fig_gals_lost} we show the mass distribution (top histogram) and star formation rate (right-hand histogram) of the classified and unclassified galaxies. We use the stellar masses and star formation rates derived from the PIPE3D. In particular, the star formation rate in each galaxy is estimated with the dust-corrected H$\alpha$ luminosity using the relationship proposed by \citet{Kennicutt_1998}. This measurement should be treated as an upper limit since the PIPE3D uses the integrated H$\alpha$ flux, which could include contamination of regions where the ionization source is not related to recent star formation events \citep[see][]{Sanchez_2022}.. We note that our classification is biased towards higher star formation rates and lower stellar masses. This is an expected (and known) caveat given that optical emission lines are sensitive to the ongoing star formation in galaxies \citep{Kewley_2008}. High-energy photons produced by young and hot stars are able to produce powerful emission lines through ionization \citep{kewley2001}. However, galaxies whose stellar population are dominated by old stars are not generally able to produce the strong emission lines needed for typical optical classifications (e.g., satisfying S/N specific criteria). Indeed, \citet{Kauffmann2003_b} finds that SDSS galaxies with higher masses present older stellar populations and that the fraction of high-mass galaxies with recent star formation is lower than in less massive galaxies. We further discuss how this impacts our AGN selection in Section \ref{sec_no_em_line}.

   We found very little or no change in the stellar mass distribution of the AGN candidates when selecting them by the different aperture steps, regardless of the used unit (kpc, arcsec, or R$_{eff}$). The typical stellar mass distribution has a median of $\sim$10$^{11}$~M$_{\odot}$. Despite the similarity in their stellar mass distributions, we do not expect an accurate AGN BPT selection when using larger apertures. This similarity can occur due to the presence of a population of galaxies with DIG-dominated regions, whose properties can mimic AGN-like signatures (these galaxies are also massive, but older; see Section \ref{sec_digs}).

   The star formation rate (derived from the integrated H$\alpha$ flux) distribution from AGN selected using small apertures is also in agreement for AGN selected at large apertures, selecting very low SFR(H$\alpha$) populations at very large apertures. Our AGN candidates (using our 2~kpc catalog) follow an offset lying mostly below the main sequence of star-forming galaxies (SFMS, see Figure \ref{fig_AGNs_MS}) and above the Retired Galaxies Sequence (RGS). Our AGN candidates are located mostly in the green valley region \citep{Salim_2014}. \cite{Leslie2016} used optical diagnostics classification and found that Seyfert-classified galaxies (with $z<0.1$) lie mostly below the SFMS. \cite{Schawinski_2009} find similar results using a sample of obscured and unobscured AGN, concluding that their AGN host galaxies ($0.01<z<0.07$) are more likely to be found in the green valley. Similar results have been discussed in a sample with higher redshifts \citep[$z<2.0$,][]{Povi_2012}, suggesting that these AGN could be part of a transitional phase, driving star-forming galaxies into a more quiescent phase. Indeed, further evidence has been found by \citet{Lefevre2019} in even higher redshift ranges ($2<z<3.8$), where strong  C~III] emitters (consistent with AGN sources) are more likely to be found below the SFMS (with an increasing outflow velocity in the strongest C~III] emitters), suggesting that negative feedback quenches star-forming galaxies into a quiescent population. We will further discuss the outflow properties in MaNGA-AGN selected galaxies in Alb\'an in prep.

\subsection{Hosts with very low S/N or no emission lines}
\label{sec_no_em_line}

   We find that the unclassified galaxies (with $S/N<3$ criteria) are typically more massive and have lower star formation rates (see Figure \ref{fig_gals_lost}) compared to galaxies with higher $S/N$. This is an expected behavior \citep[e.g.,][]{Brinchmann2004}, that happens since these galaxies are dominated by old stellar populations and have low cold gas fractions \citep{Wylezalek_2022}. For example, \citet{Brinchmann2004} found (in a sample of SDSS galaxies) that the flux of the [O~III]$\lambda$5008 emission line decreases with stellar mass, which thus typically has lower $S/N$ (consistent with the discussion in Section \ref{sec_em_line}). This trend reverses for the very high mass galaxies due to the prevalence of AGN at these stellar masses \citep[seen in][]{Kauffmann2003}.

   Using our 2~kpc aperture classification, we take the 4000~\r{AA} (D4000) break as a proxy of stellar age \citep{Kauffmann2003,Kauffmann2003_b}. While the MaNGA sample and our classified targets have a median $D_{4000}$ value of $\sim1.36$ and $\sim1.46$ respectively, our unclassified sample of galaxies is typically older, with a median of $D_{4000}\sim1.71$ and with very low EW(H$\alpha$) with a median of $\sim$0.22~\r{AA}. We also find that these galaxies typically occupy the Green Valley and Quiescent region on the D4000 vs. stellar mass diagram. Thus, we typically exclude old and high stellar mass galaxies with low star formation rates. A fraction of these discarded galaxies are classified as AGN through radio-selection methods in \citet{Comerford2020}, where they discuss the radio-mode AGN as a possible final phase in the AGN evolution.

   In a very early release of the MaNGA survey, \cite{Belfiore2016} used 646 galaxies to study the spatially resolved BPT diagrams focusing on low ionization emission-line region (LIER) galaxies. They propose a classification scheme where 151 galaxies had spaxels with very low signal-to-noise in the emission lines or no emission lines at all. They use the spaxels of all the galaxies that were not able to be classified and compute their EW(H$\alpha$), finding a median of 0.5 \r{AA}. For our unclassified sources, we also compute the median EW(H$\alpha$) from the central spaxel as well as the entire footprint, finding that the values do not exceed 0.4 \r{AA}. This median EW(H$\alpha$) is consistent with Belfiore et al., who classify line-less galaxies as the ones with EW(H$\alpha)<1.0\r{AA}$ within one effective radius. Our AGN selection excludes, by default, the galaxies whose aperture-averaged EW(H$\alpha$) is lower than 3~\r{AA}. Therefore, we do not expect a significant contribution of AGN candidates (based on our selection algorithm) coming from galaxies with no emission lines due to our $S/N$ cut.

%--------------------------------------------------- One column table
   \begin{table}
      \caption[]{MaNGA-DR17 catalog for optical diagnostics classification excluding [O~I] BPT diagram using a 2~kpc and a $3\arcsec\times 3\arcsec$ aperture.}
         \label{table1}
     $$ 
         %\begin{array}{p{0.5\linewidth}ll}
         \begin{array}{llll}
            \hline
            \noalign{\smallskip}
            Class      &  2~kpc & 3\arcsec\times 3\arcsec & Crossmatch\\
            \noalign{\smallskip}
            \hline
            \noalign{\smallskip}
            Star-forming & 3574 & 3623  &  3538  \\
            Seyfert      & 396  & 343   &  313   \\
            LINER        & 1959 & 2235  &  1824  \\
            %LINER_{EW(H\alpha)>3}    & 170 &164  &  --  \\
            Composite    & 1038 & 1085  &  972   \\
            Ambiguous    & 526  & 538   &  416   \\
            Total        & 7493 & 7824 &  7404    \\
            \hline
            AGN\tablefootmark{a}         & 419 & 399 & 378 \\
            \hline
            \noalign{\smallskip}
         \end{array}
     $$ 
     
    \tablefoot{Only around 7500 galaxies from the total MaNGA sample had available (enough $S/N$, see Section \ref{sec3}) emission-line ratios when excluding the [O~I]/H$\alpha$ BPT diagram to conduct this classification.\\
    \tablefoottext{a}{This corresponds to the final set of AGN candidates selected by the catalog when including the EW(H$\alpha$)>3 criteria to the Seyfert and LINER BPT-selected galaxies.}
    }
   \end{table}

%--------------------------------------------------- One column table
   \begin{table}
      \caption[]{MaNGA-DR17 catalog for optical diagnostics classification including the [O~I] BPT diagram using a 2~kpc and a $3\arcsec\times 3\arcsec$ aperture.}
         \label{table2}
     $$ 
         \begin{array}{llll}
            \hline
            \noalign{\smallskip}
            Class      &  2~kpc & 3''x3'' & Crossmatch\\
            \noalign{\smallskip}
            \hline
            \noalign{\smallskip}
            
            Star-forming & 2769 & 2850  &  2679   \\
            Seyfert      & 308 & 278   & 255      \\
            LINER        & 391 & 400  &  279   \\
            %Always AGN side (for 2kpc version: 1868)
            Composite    & 591  & 589   &  528    \\
            Ambiguous    & 2262  & 2582  &  1918  \\
            Total        & 6321 & 6699 &  6124    \\
            \noalign{\smallskip}
            \hline
            AGN \tablefootmark{a}  & 407 & 395 & 371 \\
            \hline
         \end{array}
     $$ 
       \tablefoottext{a}{This corresponds to the final set of AGN candidates selected by the catalog when including the EW(H$\alpha$)>3 criteria to the Seyfert and LINER BPT-selected galaxies.}
   \end{table}
%----------------------

\section{Comparison between AGN catalogs}
\label{sec_comparisons}

   To test our classification catalogs, in this section we aim to study the impact of aperture size (2~kpc and 3$''$ specifically in this section) on the final AGN classification. We also explore the reasons for the discrepancies and/or cross-matches between reported catalogs from previous literature (described in Section \ref{sec2}) and our classification described in Section \ref{sec3}. When comparing our catalogs to those from the literature, we restrict our catalogs to match the corresponding MaNGA sample (e.g., if an AGN selection was made using the MPL5 sample, we only use our AGN selected from that specific subsample).

\subsection{Internal comparison of our catalogs}

   As seen in Table \ref{table1} and Table \ref{table2} (see also Figure \ref{how_much_AGN}), the total number of classified galaxies (after the $S/N$ criteria) is greater when using a $3\arcsec\times3\arcsec$ aperture than the 2~kpc aperture. The latter is not surprising, given that more than half of the sample has a redshift which leads to the aperture of 3$\arcsec$ to correspond to physical sizes greater than 2~kpc. 
   This means that for an aperture of $3\arcsec\times3\arcsec$ in this sample, more pixels are considered, increasing the chance of meeting our $S/N$ criteria.

\begin{figure}
\centering
\includegraphics[width=\hsize]{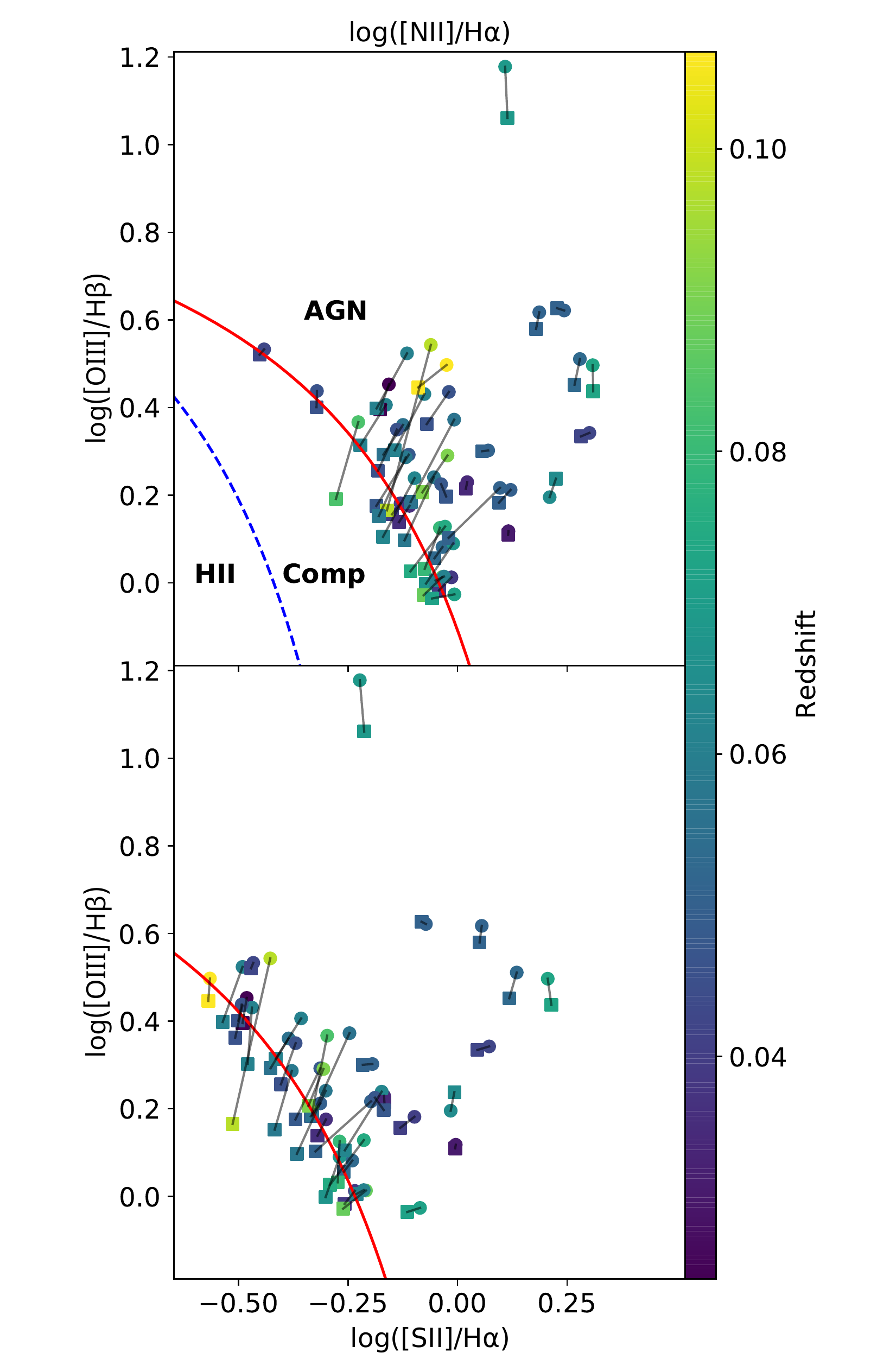}
  \caption{BPT distribution for galaxies that were selected by the 2~kpc aperture (circles) but not by the $3\arcsec\times3\arcsec$ aperture (squares); each pair is connected by a black line. Both catalogs are performed as in Section \ref{sec:optical_Select} excluding the [OI]/H$\alpha$ diagram. Each target is colored by its redshift. We plot the empirical division lines that will give each target a specific classification (e.g., AGN-like galaxy, Composite object, or HII-star forming galaxies): red lines correspond to \cite{kewley2001}, and the blue dashed line on the left plot corresponds to \cite{Kauffmann2003}.}
     \label{kpc_vs_3x3}
\end{figure}

   Even though there are more classifiable targets in the $3\arcsec\times3\arcsec$ aperture catalog, fewer targets are selected as AGN by the $3\arcsec\times3\arcsec$ aperture. Using only the [N~II]/H$\alpha$ and [S~II]/H$\alpha$ BPT diagrams, we identify 41 galaxies that are selected to be AGN based on the 2~kpc aperture but not by the $3\arcsec\times3\arcsec$ aperture. We show these galaxies in Figure \ref{kpc_vs_3x3}, revealing the impact on the BPT distribution when using the kpc vs. the arcsecond based aperture. The most common reason for this disagreement is that the (larger) $3\arcsec\times3\arcsec$ aperture pushes more targets to the star-forming regime. 
   We observe that the lowest redshift galaxies are the ones that least change the position in the diagrams as the difference in physical size becomes more relevant at higher redshifts. We find that 11 out of these 41 galaxies do not meet the EW(H$\alpha$) criterion (EW(H$\alpha$)>3\r{A}) anymore.
   
   If we reverse the comparison, we find that 21 galaxies are selected as AGN candidates based on the $3\arcsec\times3\arcsec$ aperture but not by the 2~kpc aperture. Most of them were not selected as AGN because they did not satisfy the EW(H$\alpha$) criteria: eleven of these did not satisfy the EW(H$\alpha$) cut, and the remaining ones were very close to the AGN division lines from the [N~II] and [S~II] BPT diagrams, and eight are classified as Ambiguous, one as star-forming and one as composite object in the 2~kpc aperture. Similar results are found when including the [O~I]/H$\alpha$ diagram in the selection algorithm, providing less AGN due to the typically low $S/N$ in the [O~I] emission line. We conclude that a 3$''$ based aperture is more prone to shift optically selected AGN galaxies (using a 2~kpc aperture) from the MaNGA survey towards a more star-forming appearance.
   
   To quantify the redshift dependence on the offset of the BPT classification, we focus on the [N~II] BPT diagram and measure the distance between the BPT position of both kpc and 3$\arcsec$ apertures. We name this offset parameter $\Delta (NII)$, measured as follows:
   
   \begin{equation}
   \label{eq_NII_offset}
       \Delta (NII)=\sqrt{(O3_{ap1}-O3_{ap2})^{2}+(N2_{ap1}-N2_{ap2})^{2}}
   \end{equation}

   Where $O3_{ap}=[OIII]/H\beta_{ap}$ and $N2_{ap}=[NII]/H\alpha_{ap}$, with $ap$ the type of aperture used.
   
\begin{figure}[t]
\centering
\includegraphics[width=\hsize]{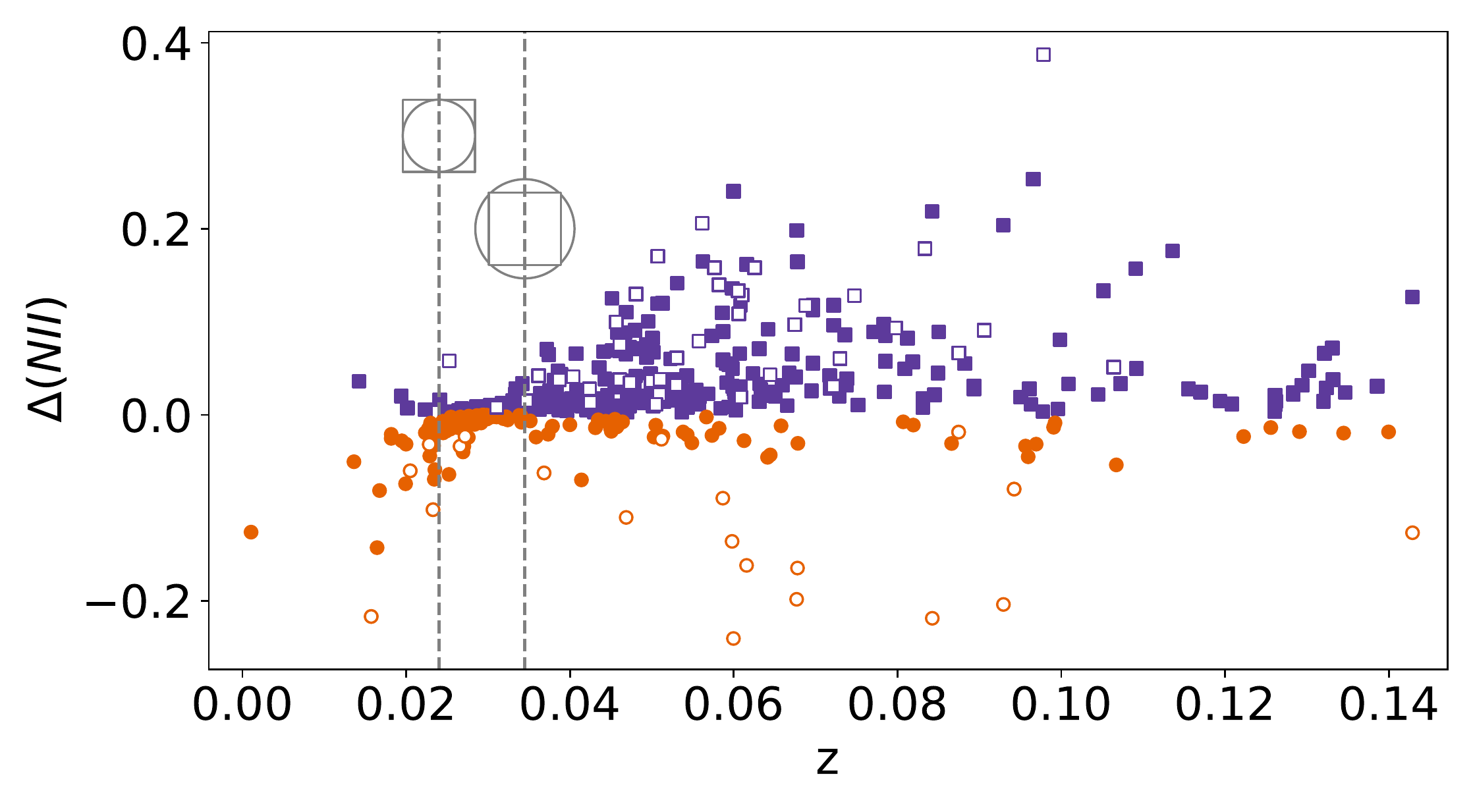}
  \caption{Magnitude of the offset between the position of a 3$\arcsec$ and 2~kpc aperture in the [NII]-BPT diagram as a function of the redshift of the target. The data displayed in this plot corresponds to the $\Delta (NII)$ from  AGN-selected galaxies using a 2~kpc aperture (and excluding the [O~I] BPT diagram). Orange solid circles correspond to galaxies whose offset slope points towards the AGN BPT region (outside the HII-BPT region). In contrast, the solid purple circles correspond to the opposite, moving galaxies towards a more star-forming appearance. Empty symbols (squares or circles) correspond to galaxies that were selected as AGN in one aperture but not in the other (e.g., the empty squares are exactly the same objects that are shown in Figure \ref{comparison_y_own}, meaning that they were selected as AGN by the 2~kpc aperture but not for the 3$\arcsec$ aperture). The dashed vertical lines correspond to the redshift values where the squared arcsecond aperture corresponds to the circumscribed and inscribed square of a circle of radius 1~kpc, respectively.}
     \label{bpt_Decrement}
\end{figure}

    We display the distribution of $\Delta (NII)$,  with $ap1=2~kpc$ and $ap2=3''$, as a function of redshift in Figure \ref{bpt_Decrement}. We highlight in purple squares the galaxies that are kicked in the direction of the HII-BPT region of the NII BPT diagram. We use negative values (see the orange circles) to show the galaxies that are shifted toward the AGN region. The latter shows that the discrepancy between flux ratio measurements (equation \ref{eq_NII_offset}) from a 2~kpc and a 3$''$ aperture in the NII-BPT diagram is more prone to increase, showing greater scatter at higher redshift. We see a greater concentration of galaxies moving towards a more HII-like region (purple squares) than galaxies shifting towards an AGN-like classification (orange circles) for these optically AGN-selected galaxies. The empty symbols (squares or circles) show the AGN candidates selected in one aperture but not in the other. The empty squares (AGN in the 2~kpc aperture but excluded by the 3$''$ aperture) reveal that more AGN candidates are being excluded by the 3$''$ aperture, supporting our findings in Figure \ref{kpc_vs_3x3}. The discrepancies in this comparison are more prominent for higher redshifted galaxies.
    
    We also see that the impact of the aperture size remains almost negligible when the physical size of the arcsecond-based aperture matches the kpc aperture size, as expected. We show in dashed lines the circumscribed and inscribed square of a circle of radius 1~kpc, displaying the redshift range where the squared aperture mostly matches the circular-kpc based aperture. Furthermore, we find no correlation between $\Delta (NII)$ and the axis ratio (b/a) when looking to our AGN candidates.
    
\subsection{Comparison to the AGN catalog from S\'anchez et al. 2018}
\label{sec_comparisons.sanchez}

\begin{figure}[t]
\centering
\includegraphics[width=\hsize]{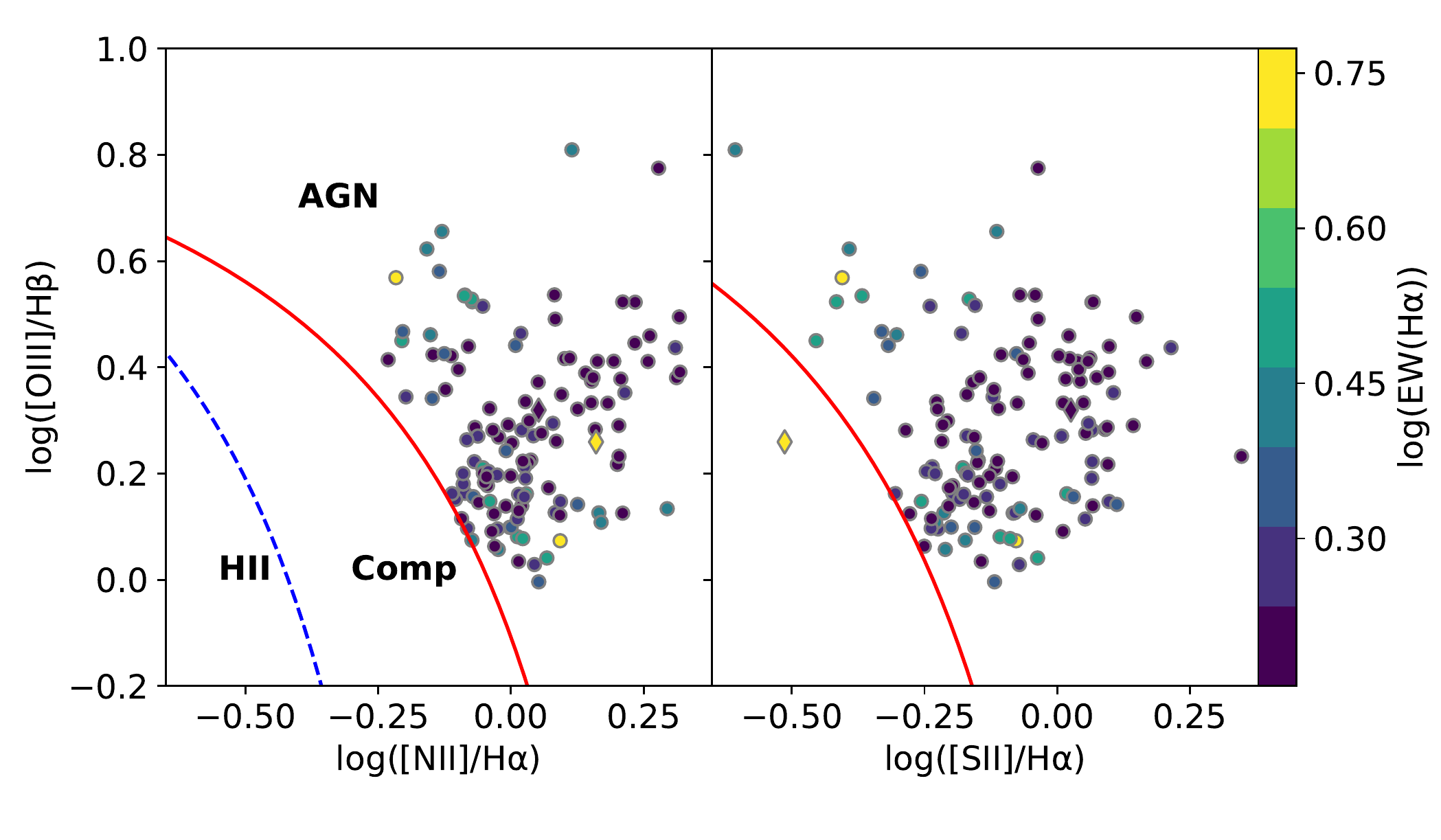}
  \caption{We display the BPT position of all the AGN candidates that are in disagreement between our 3$\arcsec$ aperture catalog (without [OI]/H$\alpha$ BPT diagram) and the catalog from \cite{sanchez2016}. Solid circles correspond to targets selected by our catalog, but not by theirs. Solid diamonds correspond to the opposite (AGN selected by them, but not by our catalog). The color on the scatter plots corresponds to the logarithmic EW(H$\alpha$). The red lines and the blue dashed line on the left plot correspond to \cite{kewley2001} and \cite{Kauffmann2003}, respectively.}
     \label{MyAGN_vs_Sanchez_general}
\end{figure}

   We compare the MaNGA AGN catalog published in \citet{sanchez2018} (which is done using the Pipe3D output; see Section \ref{mpl5sanchez}) to our $3\arcsec\times3\arcsec$ aperture catalog. To make a consistent comparison, we use the same EW(H$\alpha$) criteria (i.e., adapting to $EW(H\alpha)>1.5$~\r{A}) following the procedure in Section \ref{sec:optical_Select} and using BPT diagnostics excluding the [O~I]/H$\alpha$ diagram. Mostly all of their selected AGN are also selected in our catalog. We find two AGN candidates with differing classifications: one due to a disagreement in position on the BPT diagram, and the other not meeting the $EW(H\alpha)$ criteria (solid diamonds in Figure \ref{MyAGN_vs_Sanchez_general}). When considering the full BPT scheme, only one extra miss-match occurs, due to the lack of classifiable pixels ($S/N>3$) in the [O~I] flux (from the DAP). If we make the same comparison with our 2~kpc aperture catalog and use the full BPT scheme, five galaxies that are selected by S\'anchez et al. are not selected by our catalog. Two with no classifiable pixels in [O~I], two not satisfying the $EW(H\alpha)$ criteria, and one with an Ambiguous BPT position.
   
   However, there are more than 40 AGN candidates (if we compare to any of our four catalogs using $EW(H\alpha)>3$~\r{A}; see Section \ref{arc_kpc_catalog}) that were not selected by S\'anchez et al., and more than 100 when we relax to $EW(H\alpha)>1.5$~\r{A}. The latter is shown in Figure \ref{MyAGN_vs_Sanchez_general}, where we show our emission line ratio measurements and adapt to their $EW(H\alpha)>1.5$~\r{A} criteria. A visual inspection suggests the fact that many AGN candidates from our catalog that are not selected by \cite{sanchez2018} are relatively close to the \cite{kewley2001} demarcation line (see Figure \ref{MyAGN_vs_Sanchez_general}). Discrepancies for these targets can be related to the quality cut that we make at $S/N>3$ for each emission line measured by the DAP, which excludes that specific pixel from our procedure, having slight flux ratio changes from the ones measured by S\'anchez et al. However, \cite{sanchez2018} reported 302 galaxies above the AGN division line \citep{kewley2001,Kauffmann2003}, suggesting that the discrepancies for the galaxies far away from the demarcation lines on the AGN selection are mostly related to the $EW(H\alpha)$ measurements. These discrepancies are a consequence of a different treatment of the stellar continuum when measuring $EW(H\alpha)$ \citep[already reported in][]{thomas2013,2018Wylezalek} and possibly due to the difference in the $S/N$ cut criteria.

   \begin{figure}[t]
\centering
\includegraphics[width=\hsize]{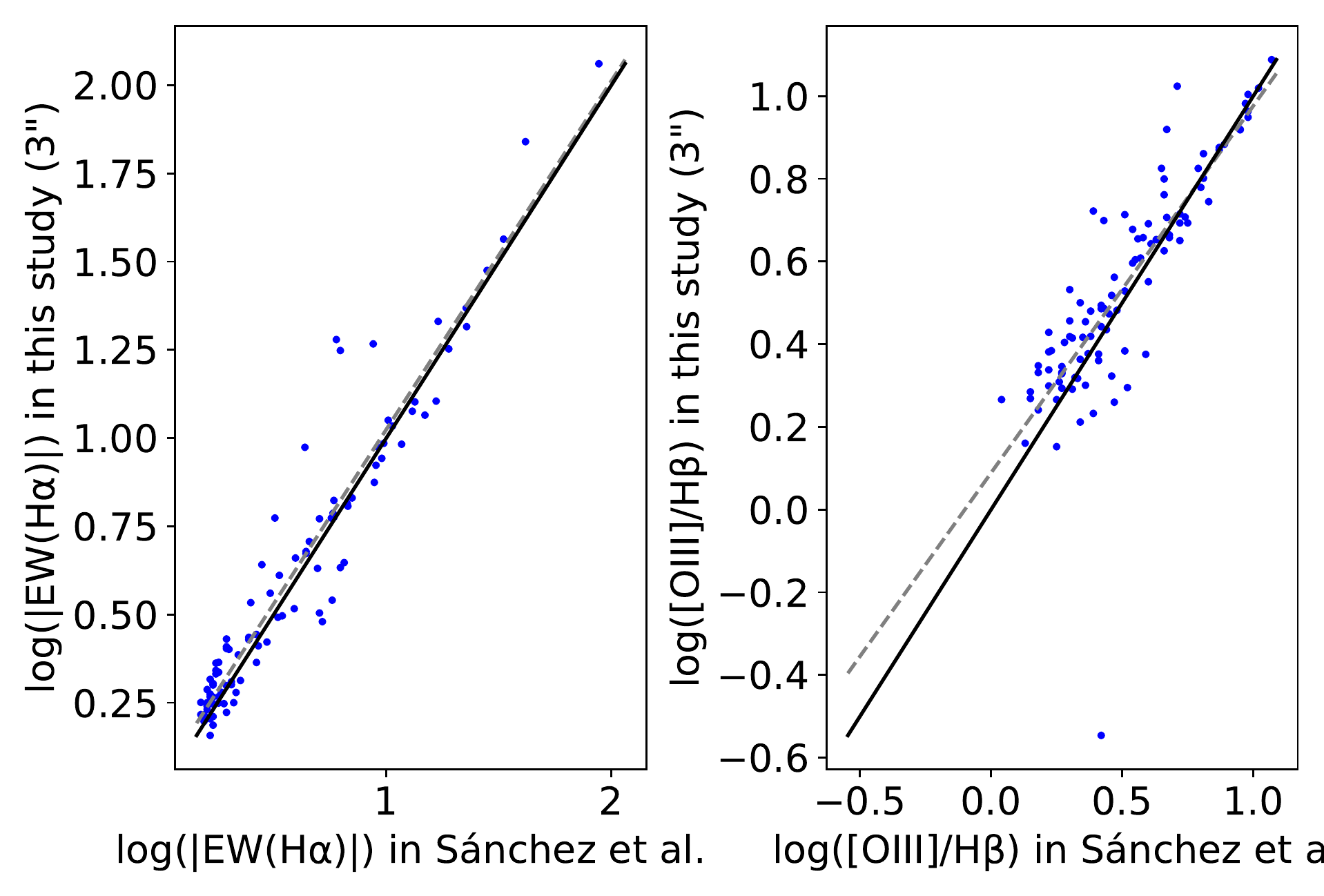}
  \caption{Blue circles correspond to a comparison between measurements of EW(H$\alpha$) values (left-hand plot) and [OIII]/H$\beta$ ratios (right-hand plot) from this work (y-axis; using a 3$\arcsec$ aperture) and Sánchez et al. (x-axis) in a logarithmic scale. The measurements shown here correspond to the AGN-candidate subsample from S\'anchez et al. The solid black line in both plots shows the one-to-one ratio that the values should follow if they had the same value. The gray dashed line in the plots shows a 1-D polynomial fit to the distribution.}
     \label{comparison_sanchez_mine_measurements_fig}

\end{figure}

   In Figure \ref{comparison_sanchez_mine_measurements_fig}, we contrast the $EW(H\alpha)$ measurements from both AGN catalogs (left plot). A 1-D polynomial suggests that the $EW(H\alpha)$ measurements are mostly in agreement. However, it can be seen that the one-to-one comparison of the $EW(H\alpha)$ values shows an important scatter. At lower $EW(H\alpha)$s, our measurements (from our 3$\arcsec$ procedure) predict typically larger values than those measured by S\'anchez et al. This supports the idea that the main discrepancy in the AGN selection comes from the differences between the $EW(H\alpha)$ values. In the right-hand plot, we compare the [O~III]/H$\alpha$ ratio measurements. In our catalog, [O~III]/H$\alpha$ has typically greater values \citep[also reported in ][when comparing the DAP measurements to Pipe3D, due to a different choice of stellar continuum treatments]{Belfiore2019}. The latter is expected since there is a better agreement in the AGN selection before applying the $EW(H\alpha)$ criteria.
   
   A minor extra consideration that could drive the discrepancy is that we are weighting pixels according to their enclosed fraction with their selected aperture (see Section \ref{sec:optical_Select}), which we do not expect to have a significant impact. We find similar results when using our 2~kpc aperture and the $EW(H\alpha)=1.5$~\r{A} criteria.

\subsection{Comparison to the AGN catalog from Rembold et al. 2017}
\label{sec_comparisons.rembold}

\begin{figure}[t]
\centering
\includegraphics[width=\hsize]{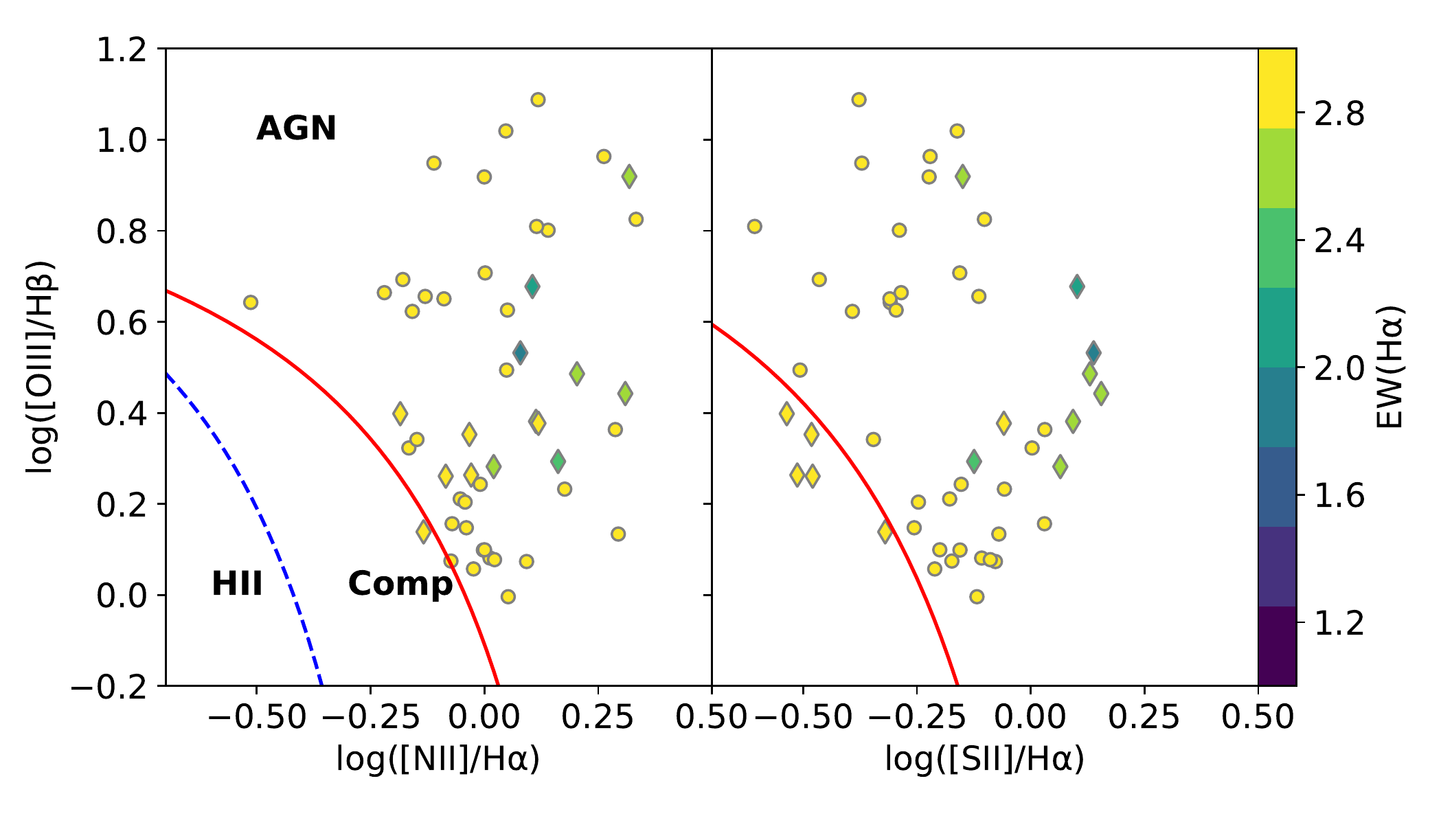}
  \caption{BPT distribution of all the AGN candidates that are in disagreement between our 3$\arcsec$ aperture and BPT excluding [OI]/H$\alpha$ diagram catalog and the catalog from \cite{rembold2017}. Solid circles correspond to targets selected by our catalog, but not by theirs. Solid diamonds correspond to the opposite (selected by them, but not by our catalog). The color on the scatter plots corresponds to a specific EW(H$\alpha$) range width whose contrast is forced between $1.0$~\r{AA}~$> EW(H\alpha)>3.0$~\r{AA} to emphasize which targets satisfied the $EW(H\alpha)>3.0$~\r{AA} condition. The red lines correspond to \cite{kewley2001} and the blue dashed line on the left plot corresponds to \cite{Kauffmann2003}.}
     \label{MyAGN_vs_Rembold}
\end{figure}

   Using our $3\arcsec\times3\arcsec$ aperture and excluding the [O~I]/H$\alpha$ BPT diagram we find that only one target that does not crossmatch with the AGN candidates from \citet{rembold2017}. However, 14 galaxies drop from the crossmatch when we adapt to the $EW(H\alpha) > 3$~\r{A} criteria. We display the miss-matched targets in Figure \ref{MyAGN_vs_Rembold} and we force the contrast of the $EW(H\alpha)$ color-bar to have its maximum value at $EW(H\alpha)=3$ (in logarithmic scale) to better distinguish why such galaxy was not classified as AGN in our procedure in terms of the EW(H$\alpha$). Furthermore, if we consider the full BPT scheme (including the [O~I]/H$\alpha$ BPT), the latter behavior holds. When we perform the same comparison with our absolute 2~kpc aperture, we find one more crossmatched target. When reversing the comparison, around 36 targets were selected by our catalogs but not by the Rembold catalog.

    Including the [OI]-based BPT diagnostic or changing the aperture (between $3\arcsec\times3\arcsec$ arcsec or 2~kpc) does not seem to impact the number of AGN-selected sources, as also found in the comparison with the S\'anchez et al. classification. Five targets are not classified as AGN candidates (but as Ambiguous) since they fall into the star-forming regime in the [S~II/H$\alpha$] diagram. Most of the extra AGN candidates from \citep{rembold2017} do not satisfy the EW(H$\alpha$) > 3~\r{AA} criteria. Specifically, the 14 and the further 36 discrepancies (see Figure \ref{MyAGN_vs_Rembold}) suggest that a $S/N$ cut plays an important role for the EW(H$\alpha$) criteria since we use a $S/N>3$. Low $S/N$ spaxels could be biased by low EW(H$\alpha$) values as this quantity is measured by the ratio between the emission-line and continuum fluxes \citep{thomas2013}. An important discrepancy comes from the difference in the stellar subtraction from MaStar (MaNGA Stellar Library) when measuring emission lines \citep{Yan_2019}. A smaller impact might be related to the way in which we measure the flux (see Figure \ref{fig_pix_weight}).
    
\subsection{Comparison to the AGN catalog from Wylezalek et al. 2018}
\label{sec_comparisons.wylezalek}

    An interesting result from the comparisons conducted above is that as we increase the aperture (either using a kpc or arcsecond aperture), our catalogs have fewer AGN in common compared to the catalogs of \citep{sanchez2018} and \cite{rembold2017}. However, the agreement between our catalogs and the one from \cite{2018Wylezalek} remains almost constant as we increase the aperture for the selection, with a small increase at greater apertures. 
    \begin{figure}[t]
\centering
\includegraphics[width=\hsize]{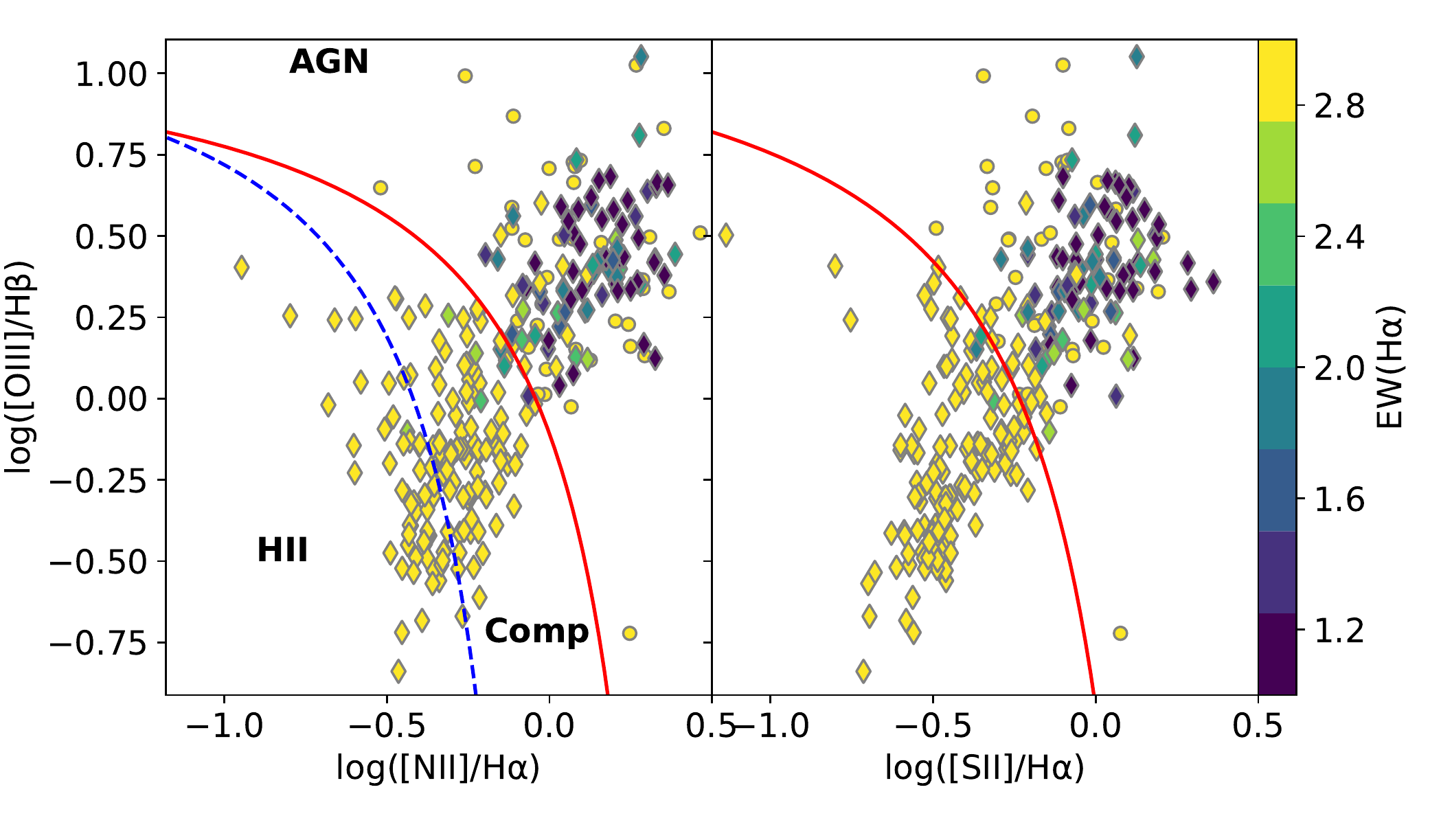}

  \caption{AGN candidates that are in the \cite{2018Wylezalek} catalog but are excluded by our AGN selection (solid diamonds) using a 2~kpc aperture without the [O~I]/H$\alpha$ BPT criteria with $EW(H\alpha)>3$~\r{A} (with the contrast forced as in Figure \ref{MyAGN_vs_Rembold}). The ones that were selected by our procedure but not by Wylezalek et al. are shown in filled circles. The dashed and solid lines have the same interpretation as described in Figure \ref{kpc_vs_3x3}.}
     \label{MyAGN_vs_Wylezalek}
\end{figure}
    
    In contrast to the previous comparisons, we find a greater disagreement between our AGN candidates and the ones from \citep{2018Wylezalek}. In \citet{2018Wylezalek}, the authors search for AGN signatures at all galactocentric distances. If we use our 2~kpc diameter aperture catalog (excluding [O~I]/H$\alpha$), we find only 56 targets in agreement (Figure \ref{MyAGN_vs_Wylezalek}). Changing between a 2~kpc or a 3$''$ aperture and the full BPT scheme has no relevant impact for this comparison. The discrepancies reported in this comparison are mostly related to the significant differences in the selection methods. Their method does not use defined apertures, but pixel fractions (e.g., if a relevant fraction of pixels is classified as AGN); every pixel is used, whether it is near or far from the center of the galaxy. \citet{2018Wylezalek} argue that using the full power of spatially resolved spectra reveals possibly hidden and/or weaker AGN candidates that may have been missed due to obscuring effects, recently turned-off AGN or AGN that are not located in the reported center of the MaNGA NSA catalog (due to mergers or off-set AGN).

\subsection{Comparison to the multi-wavelength AGN catalog from Comerford et al. 2020}

\begin{figure}[t]
\centering
\includegraphics[width=\hsize]{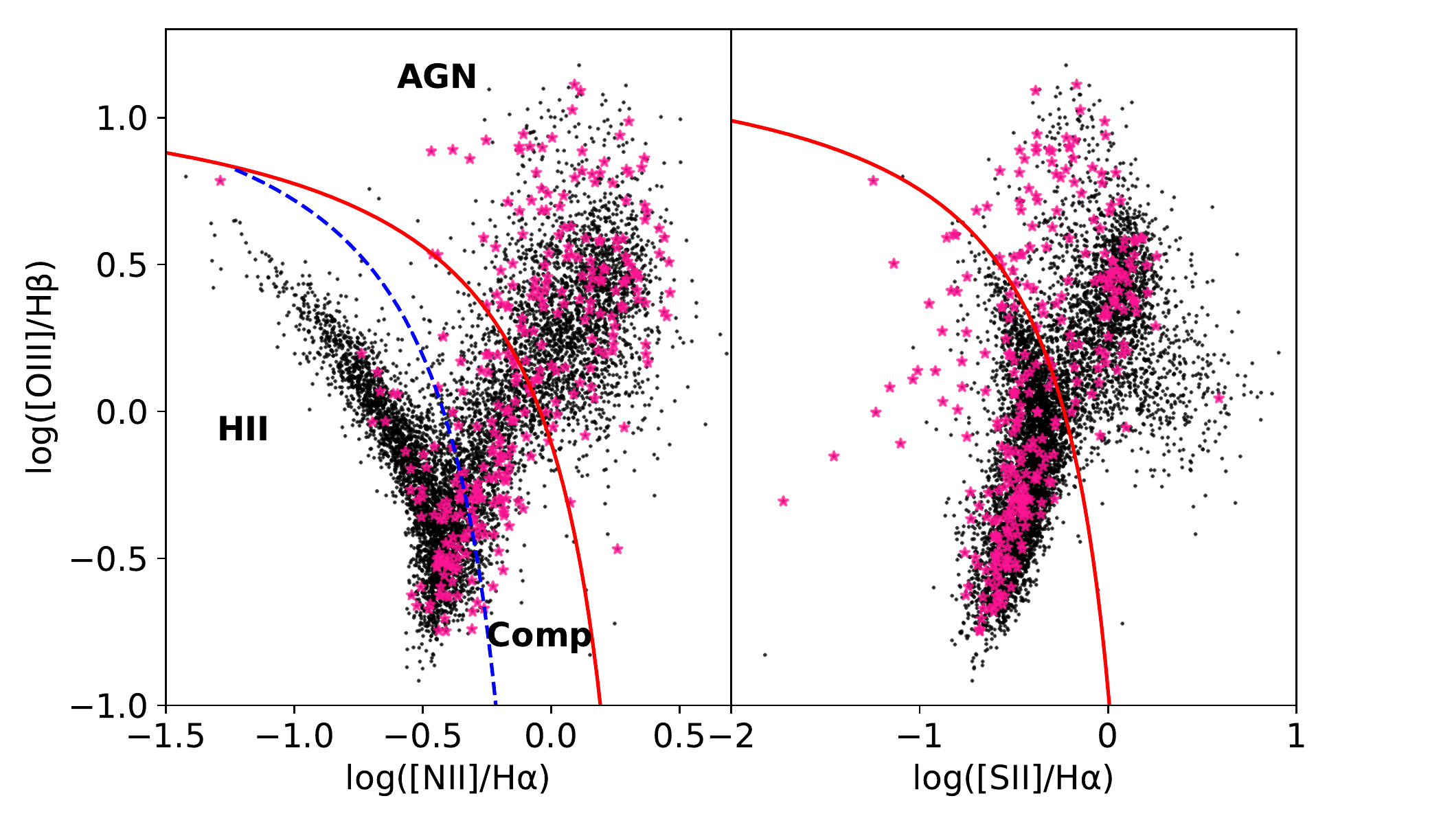}
  \caption{We show our MaNGA BPT classification using a 2~kpc aperture. We highlight the multi-wavelength (based on radio, infrared, broad-line, and X-ray selection methods, see Section \ref{Comerford_catalog}) selected AGN candidates from \cite{Comerford2020} with pink-colored star-like symbols.}
     \label{Fig4}
\end{figure}

   From their 406 AGN-selected galaxies, we find only 73 and 69 galaxies in agreement from our 2~kpc and 3$''$ (excluding [O~I]/H$\alpha$) AGN catalog, respectively when compared to the multi-wavelength AGN catalog from \citet{Comerford2020} (for a description of this catalog, see Section \ref{Comerford_catalog}). These numbers decrease as we increase the size of our apertures. A better agreement is found when we relax the $EW(H\alpha)$ criteria. Around 100 galaxies that were classified as AGN in \citet{Comerford2020} were not able to be classified by our 2~kpc aperture catalog due to our $S/N$ cut. In Figure \ref{Fig4} we display the BPT diagrams highlighting the non-optical classified AGN from the \citet{Comerford2020} catalog. The AGN sample of \citet{Comerford2020} seems to not show any preference along with our BPT classification regime. This highlights the fact that not all AGN can be classified using optical emission line diagnostics, revealing the known caveats of BPT selection.

\section{Conclusions}
\label{conclusions}

   We have used the final data release (DR17) of the SDSS-MaNGA survey (10,010 galaxies) to classify galaxies based on their BPT diagnostics. We have created catalogs using a range of apertures (in units of kpc, arcsecond and effective radius) that we base our classification on. In each aperture, we measure the BPT diagnostics and include an $EW(H\alpha)$ cut for the final AGN classification. We have studied how galaxies can change their classification as we increase the aperture used for their characterization. 
   Our main results are as follows:
   
      \begin{itemize}
      \item MaNGA AGN candidates for small ($< 6$~kpc) kpc-based apertures lie below the main sequence of star-forming galaxies (see Figure \ref{fig_AGNs_MS}), in agreement with previous results \citep[e.g.][]{sanchez2018,Comerford2020}. These AGN candidates are massive with stellar of around $\sim10^{11}M_{\odot}$ and a median H$\alpha$-derived star formation rate of $\sim1.44$~$M_{\odot}~y^{-1}$.
      
      \item We show that the number of galaxies classified as AGN decreases with increasing aperture size up to $\sim$ 6~kpc, 6$''$ or $1.2~R_{eff}$, depending on the units used, respectively. We argue that this is due to the compact nature of the AGN phenomenon. As the aperture size increases, one gets closer to capturing the entire spectrum of a galaxy, potentially reducing the dominance of AGN emission and positioning the observed galaxies towards the Composite or Star-forming region.
      
      \item Intriguingly, at apertures greater than 6~kpc, 6$\arcsec$ or $1.2~R_{eff}$), the trend reverses and the number of galaxies classified as AGN starts to increase (see Figure \ref{how_much_AGN}). The stacked radial cumulative surface brightness profiles for the AGN candidates that are only found in very large apertures (but not in the smaller ones) show very low H$\alpha$ surface brightness (see Figure \ref{radial_surf_dig_effect}). This is consistent with AGN selection being strongly contaminated by the effect of DIG-dominated regions. This implies that using a very large aperture selects fewer true AGN.
            
      \item When we compare our catalog with MaNGA AGN aperture-based catalogs from the literature (using previous data releases) we find the following important parameter. In addition to the classical BPT line ratios, the measurement of the $EW(H\alpha)$ is a commonly used and important additional diagnostic. In our comparisons, we show that different treatments to fit the stellar continuum have a strong impact on the faint AGN population. This is because EW(H$\alpha$) measurements are sensitive to small changes of the continuum model. An AGN selection based on optical diagnostics can therefore vary greatly between different analyses (see Figures \ref{comparison_sanchez_mine_measurements_fig}, \ref{MyAGN_vs_Sanchez_general}, and \ref{MyAGN_vs_Rembold}). Additional discrepancies on the selection of AGN candidates are related to the strong differences in the selection criteria (see Section \ref{wylezalek_catalog} and Figure \ref{MyAGN_vs_Wylezalek}) and the choice of the observed wavelength range when classifying galaxies (see Section \ref{Comerford_catalog} and Figure \ref{Fig4}).

   \end{itemize}

   With this work, we show that the choice of aperture size impacts optical AGN selection and galaxy classification. In comparison to single fibre optical galaxy classification, IFU observations offer a more complete characterization of the origin of ionizing sources in galaxies (e.g., identifying DIG regions)  and allow one to minimize aperture effects (e.g., using a redshift-independent aperture), therefore, reducing the bias in AGN selection techniques.

%All comments were already taken into account

\begin{acknowledgements}
      D.W. acknowledges support through an Emmy Noether Grant of the German Research Foundation, a stipend by the Daimler and Benz Foundation and a Verbundforschung grant by the German Space Agency.\\

      Funding for the Sloan Digital Sky Survey IV has been provided by the Alfred P. Sloan Foundation, the U.S. Department of Energy Office of Science, and the Participating Institutions. SDSS-IV acknowledges support and resources from the Center for HighPerformance Computing at the University of Utah. The SDSS web site is www.sdss.org.\\

      SDSS-IV is managed by the Astrophysical Research Consortium for the Participating Institutions of the SDSS Collaboration including the Brazilian Participation Group, the Carnegie Institution for Science, Carnegie Mellon University, the Chilean Participation Group, the French Participation Group, Harvard-Smithsonian Center for Astrophysics, Instituto de Astrof\'isica de Canarias, The Johns Hopkins University, Kavli Institute for the Physics and Mathematics of the Universe (IPMU) / University of Tokyo, the Korean Participation Group, Lawrence Berkeley National Laboratory, Leibniz Institut f\"ur Astrophysik Potsdam (AIP), Max-Planck-Institut f\"ur Astronomie (MPIA Heidelberg), Max-Planck-Institut f\"ur Astrophysik (MPA Garching), Max-Planck-Institut f\"ur Extraterrestrische Physik (MPE), National Astronomical Observatories of China, New Mexico State University, New York University, University of Notre Dame, Observatario Nacional / MCTI, The Ohio State University, Pennsylvania State University, Shanghai Astronomical Observatory, United Kingdom Participation Group, Universidad Nacional Autonoma de M\'exico, University of Arizona, University of Colorado Boulder, University of Oxford, University of Portsmouth, University of Utah, University of Virginia, University
\end{acknowledgements}

\bibliographystyle{aa}
\bibliography{bibliography}

\end{document}